\documentclass[aps, pra, reprint, superscriptaddress]{revtex4-1}

\usepackage{txfonts}

\usepackage[pdftex]{graphicx}
\usepackage{dcolumn}% Align table columns on decimal point
\usepackage{bm}% bold math
\usepackage{soul}
\usepackage{hyperref}
\hypersetup{
    colorlinks=true,        % false: boxed links; true: colored links
    linkcolor=blue,          % color of internal links (change box color with linkbordercolor)
    citecolor=blue,        % color of links to bibliography
    filecolor=magenta,      % color of file links
    urlcolor=blue           % color of external links
}
\usepackage{color}

\begin{document}

\title{Magnetic-field-induced spin crossover of Y-doped Pr$_{0.7}$Ca$_{0.3}$CoO$_{3}$}

\author{Akihiko~Ikeda}
\email[E-mail: ]{ikeda@issp.u-tokyo.ac.jp}
\author{Suyeon~Lee}
\author{Taku~T.~Terashima}
\author{Yasuhiro~H.~Matsuda}
\email[E-mail: ]{ymatsuda@issp.u-tokyo.ac.jp}
\author{Masashi~Tokunaga}
\affiliation{Institute for Solid State Physics, University of Tokyo, Kashiwa, Chiba 277-8581, Japan}
\author{Tomoyuki~Naito}
\affiliation{Faculty of Science and Engineering, Iwate University, Morioka, Iwate 020-8551, Japan}

\date{\today}

\begin{abstract}

The family of hole-doped Pr-based perovskite cobaltites, Pr$_{0.5}$Ca$_{0.5}$CoO$_{3}$ and (Pr$_{1-y}$RE$_{y}$)$_{0.3}$Ca$_{0.7}$CoO$_{3}$ (where RE is rare earth) has recently been found to exhibit simultaneous metal-insulator, spin-state, and valence transitions.
We have investigated magnetic-field-induced phase transitions of (Pr$_{1-y}$Y$_{y}$)$_{0.7}$Ca$_{0.3}$CoO$_{3}$ by means of magnetization measurements at 4.2$-$100 K up to an ultrahigh magnetic field of 140 T with the chemical pressure varied by $y$ = 0.0625, 0.075, 0.1.
The observed magnetic-field-induced transitions were found to occur simultaneously with the metal-insulator transitions up to 100 T.
The obtained magnetic field-temperature ($B$-$T$) phase diagram and magnetization curves are well analyzed by a spin-crossover model of a single ion with interion interactions.
On the other hand, the chemical pressure dependence of the experimentally obtained magnetization change during the phase transition disagrees with the single ion model when approaching low temperatures.
The significant $y$ dependence of the magnetization change at low temperatures may arise from the itinerant magnetism of Co$^{3+}$ in the paramagnetic metallic phase, where the chemical pressure enhances the exchange splitting by promoting the double-exchange interaction.
The observed $B$-$T$ phase diagrams of (Pr$_{1-y}$Y$_{y}$)$_{0.7}$Ca$_{0.3}$CoO$_{3}$ are quite contrary to that of LaCoO$_{3}$, indicating that in (Pr$_{1-y}$Y$_{y}$)$_{0.7}$Ca$_{0.3}$CoO$_{3}$ the high-field phase possesses higher entropy than the low-field phase, whereas it is the other way around in LaCoO$_{3}$.

\end{abstract}

%\pacs{75.30.Wx, 75.25.Dk, 75.47.Lx, 75.30.Cr}
%\keywords{}

\maketitle


\section{Introduction}

Strong correlations among electrons in transition-metal oxides often lead to the coupling of multiple degrees of freedom in solids, such as charge, orbital, and spin, that give rise to exotic phenomena such as superconductivity, colossal magnetoresistance, metal-insulator transition, multiferroics, and so forth.
Cobaltites are considered unique among transition-metal oxides for their spin-state degrees of freedom.
One of the most interesting phenomena expected in cobaltites is the ordering of spin states.
A perovskite cobaltite, LaCoO$_{3}$, has been studied for over half a century.
The spin states of octahedrally coordinated Co$^{3+}$ ($3d^{6}$) are classified into low-spin (LS: $t_{2g}^{6}e_{g}^{0}$), intermediated-spin (IS: $t_{2g}^{5}e_{g}^{1}$), and high-spin (HS: $t_{2g}^{4}e_{g}^{2}$) states according to their total spin angular momentum $S=0$, 1, 2, respectively.
Whereas the nonmagnetic and insulating ground state at $T< 30$ K in LaCoO$_{3}$ has been identified as the LS state with neutron scattering \cite{Asai1994, Saitoh}, its electronic excited state is still controversial.
Various electronic energy schemes and ordered phases have been proposed, represented by the HS/LS order \cite{Raccha} and the uniform IS state with orbital order \cite{Korotin}. 
However, no such order or short-range correlations have been identified concretely by microscopic measures.

On the other hand, a hole-doped Pr-based perovskite cobaltite, Pr$_{0.5}$Ca$_{0.5}$CoO$_{3}$, was found to undergo the first-order and simultaneous magnetic, metal-insulator and valence transitions \cite{Tsubouchi2002}.
The phase transition is considered analogous to the virtual phase change \cite{Knizek2010} between the ferromagnetic metallic phase of La$_{1-x}$Sr$_{x}$CoO$_{3}$ with $x>0.3$ \cite{Saitoh1} and the diamagnetic insulator phase of LaCoO$_{3}$ at low temperatures \cite{Asai1994}.
The pressure and chemical pressure effects were found to be significant in Pr$_{1-x}$Ca$_{x}$CoO$_{3}$ with a small hole concentration ($x<0.5$) \cite{Fujita, Tsubouchi2004, Naito2010}.
The pressure suppresses the ferromagnetic order, and the paramagnetic insulator phase emerges \cite{Fujita,  Phelan2014}.
The phase transition was claimed to be the spin-state transition between the Co$^{3+}$ insulating LS state and metallic IS state \cite{Tsubouchi2002, Tsubouchi2004}.
Recent theoretical analyses of the two-orbital Hubbard model indicate the Bose-Einstein condensation of excitons is a possible origin of the insulating ground state of the Pr-based cobaltites \cite{Kunes1, Kaneko, Nasu}.

In the last decade, studies employing high magnetic fields have revealed nontrivial field induced phases of perovskite cobaltites.
As for LaCoO$_{3}$, sharp first-order magnetic phase transitions accompanied by large lattice expansions at around 60$-$65 T \cite{Sato2009, Moaz, Rotter}.
An even wider $B$-$T$ phase diagram for LaCoO$_{3}$ has been constructed at high temperature up to 120 K and high magnetic fields up to 135 T \cite{Ikeda}.
The phase diagram revealed the two low entropy phases emerge in the high-field region, which is counterintuitive to the single-ion picture.
Rather, it indicates the presence of the inherent strong correlation between LS and HS or IS, which is debated theoretically \cite{Sotnikov, Tatsuno}.
On the other hand, for (Pr$_{1-y}$Y$_{y}$)$_{0.7}$Ca$_{0.3}$CoO$_{3}$, the $B$-$T$ phase diagram has been revealed by magnetization and conductivity measurements for $y=0.0625$ using static magnetic fields \cite{Naito2014} and conductivity measurements for $y=0.075$ and 0.1 using ultrahigh magnetic fields \cite{Suyeon}, where Y acts as the chemical pressure.
Those studies commonly revealed the field-induced insulator-metal transitions.
The transition fields decrease with increasing temperature, which is contrary to the case of LaCoO$_{3}$ \cite{Ikeda} and more familiar among the spin-crossover systems in the single-ion picture \cite{Kimura2005, Kimura2008}.

In this paper, we present a high-field magnetization study of (Pr$_{1-y}$Y$_{y}$)$_{0.7}$Ca$_{0.3}$CoO$_{3}$ in a wide temperature range from 4.2 to 100 K and up to ultrahigh magnetic fields of 140 T, and the chemical pressure dependence is varied as $y=0.0625$, 0.75, 0.1.
A series of field-induced magnetic transitions were observed.
The obtained $B$-$T$ phase diagrams are quite contrary to that of LaCoO$_{3}$, which is well analyzed with the spin-crossover model of a single ion from LS to IS induced by magnetic field.
On the other hand, the amount of the magnetization change at the phase transitions was found to be too strongly dependent on the chemical pressure $y$ to be explained by the above model.
The possible origins of the discrepancy are discussed in terms of the itinerant magnetism of Co$^{3+}$ in the IS state.
Further, the vanishingly small latent heat at $y =0.0625$ may suggest the presence of a critical point at low temperatures.

\section{Experiment}
Magnetization measurements (Pr$_{1-y}$Y$_{y}$)$_{0.7}$Ca$_{0.3}$CoO$_{3}$ for $y=0.0625$,  0.075, 0.1 were carried out as follows. Polycrystalline samples of (Pr$_{1-y}$Y$_{y}$)$_{0.7}$Ca$_{0.3}$CoO$_{3}$ for $y=0.0625$,  0.075, 0.1 were used whose transition temperatures at zero field are $T_{\mathrm{C}}=$ 42, 64, 96 K, respectively \cite{Naito2010}.
A non-destructive pulse magnet at the Institute for Solid State Physics, University of Tokyo, was used to generate pulsed magnetic fields of up to 50 T with 37 ms duration for the magnetization measurements using the induction method.
The magnetization probe of the non-destructive pulse magnet was calibrated to the absolute value.
Further, a horizontal type single-turn coil (H-STC), a destructive pulse magnet in the Institute for Solid State Physics, University of Tokyo, was also used to generate the magnetic fields of up to 140 T with 6 $\mu$s duration also for the magnetization measurements using the induction method.
In the case of the H-STC, the magnetization pickup coil consists of a pair of counter-wound coils of 20 turns, either of which holds the sample inside \cite{Takeyama2012}.
The induction voltage $V_{M}\propto dM/dt$ was recorded with a digital oscilloscope.
A He flow-type cryostat made of Bakelite was used to cool the sample down to 4.2 K.
The temperature was monitored using a chromel-constantan thermocouple and a RuO$_{2}$ thermometer \cite{Amaya, Ikeda}.
The induction voltage of the magnetic field $V_{B} \propto dB/dt$ was monitored at the position adjacent to the sample with a magnetic field pick-up coil calibrated to the absolute value with a precision of $\sim 1$\%.

\begin{figure}[b]
\begin{center}
\includegraphics[angle=0, scale=0.6, clip]{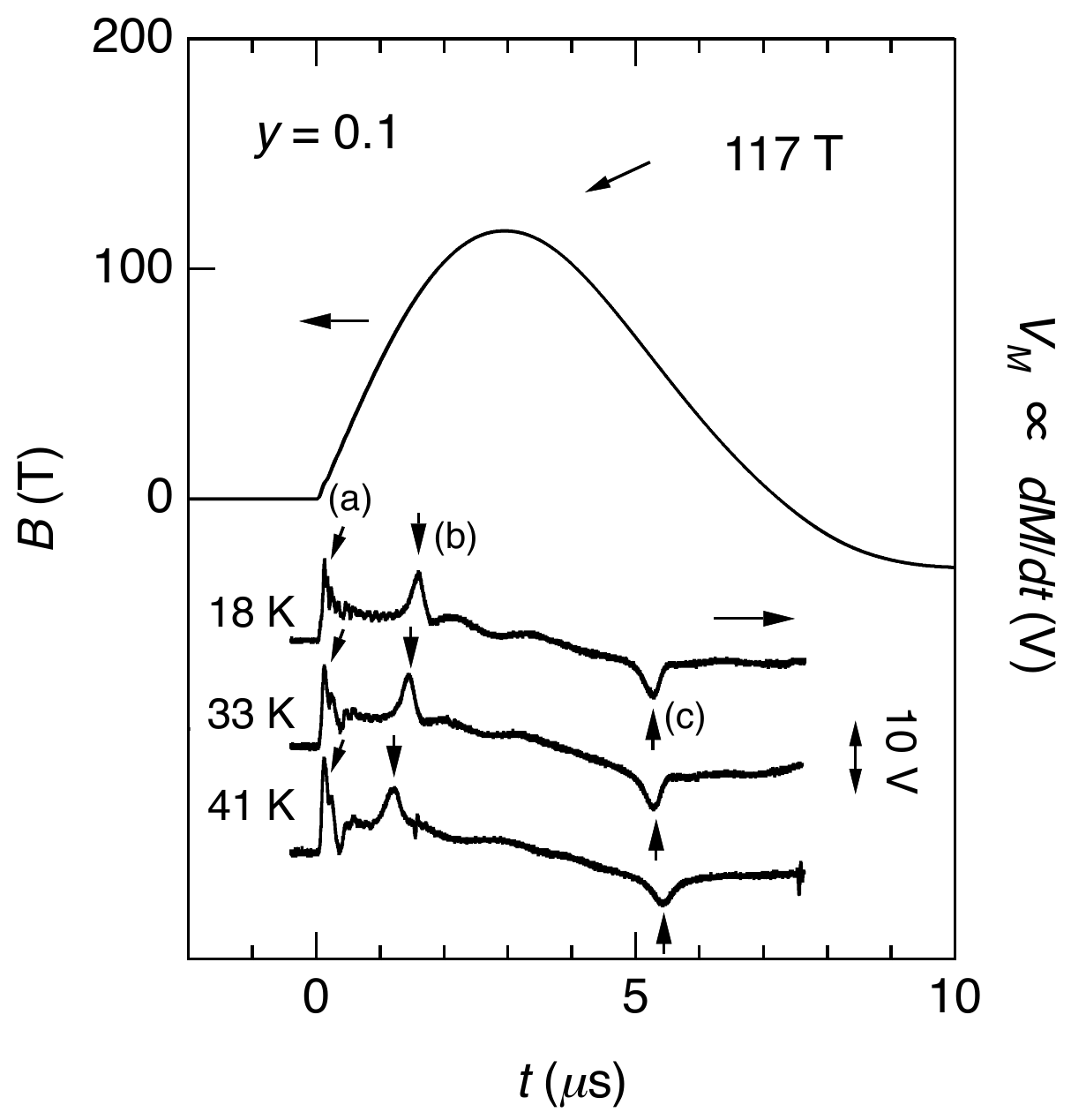} 
\caption{Representative data of the induction voltage of the magnetization of (Pr$_{1-y}$Y$_{y}$)$_{0.7}$Ca$_{0.3}$CoO$_{3}$ ($y=0.1$) and the pulsed magnetic field generated using the horizontal single-turn coils.\label{fig1}}
\end{center}
\end{figure}

\begin{figure*}
\begin{center}
\includegraphics[angle=0, scale=1, clip]{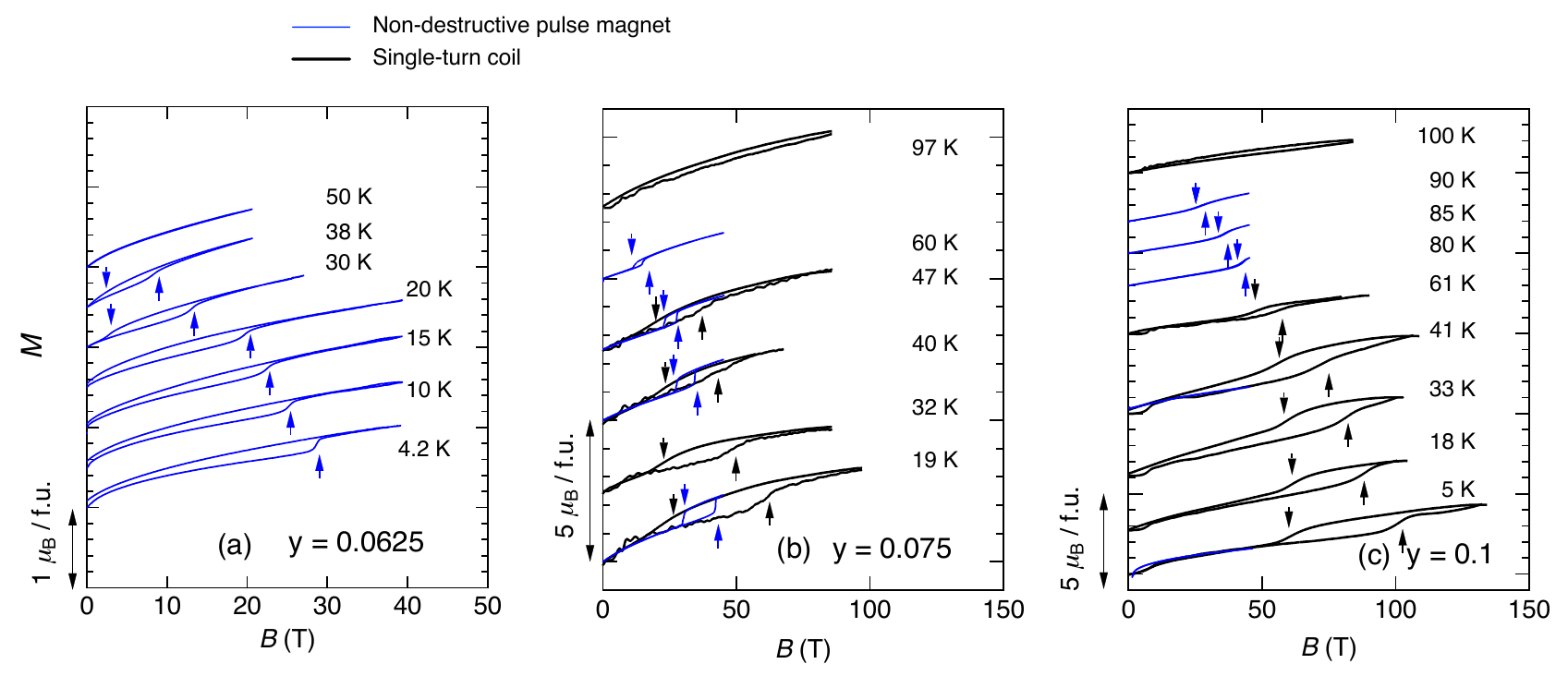} 
\caption{Magnetization curves of (Pr$_{1-y}$Y$_{y}$)$_{0.7}$Ca$_{0.3}$CoO$_{3}$ for $y=$ (a) 0.0625, (b) 0.075, and (c) 0.1, measured at various temperatures using the induction method. \label{fig2}}
\end{center}
\end{figure*}

\section{Result}

Representative results of the measured $V_{M}$ of (Pr$_{1-y}$Y$_{y}$)$_{0.7}$Ca$_{0.3}$CoO$_{3}$ with $y=0.1$ at 18, 33, and 41 K and pulsed magnetic fields are shown in Fig. \ref{fig1}.
First, the peaks in Fig. \ref{fig1} indicated by arrows (b) and (c) indicate the steep increase and decrease of $M$ in the ascending and descending magnetic fields, respectively.
The peak positions and sharpness of the peaks in $V_{M}$ are temperature dependent.
Second, we note that the sharp peaks just after $t = 0$~$\mu$s indicated by arrow (a) in Fig. \ref{fig1} are in good agreement with the observed magnetization increase at below 1 T \cite{Hejtmanek2013}, which may originate in the Curie paramagnetism of Pr$^{4+}$ or Co$^{4+}$, whose disappearance in the descending fields may be due to the heating effect.
Third, each $dM/dt$ curve commonly has a component proportional to $dB/dt$ which shows little temperature dependence.
A series of magnetization curves of (Pr$_{1-y}$Y$_{y}$)$_{0.7}$Ca$_{0.3}$CoO$_{3}$ is obtained by integrating $V_{M}$ and is plotted for $y=0.0625, 0.075, 0.1$ in Figs. \ref{fig2}(a)-\ref{fig2}(c) as the thick black curves .
Arrows pointing upwards and downwards indicate the magnetic transition fields in the ascending and descending fields, respectively.
The absolute values for $M$ obtained with the H-STC experiments are evaluated by fitting to the magnetization curves obtained using non-destructive pulses up to 50 T.
As pointed out for the $dM/dt$ curves in Fig. \ref{fig1}, we observed three components in the magnetization curves in Fig. \ref{fig2}.
First, the sharp magnetic transitions are quite dependent on temperature and chemical pressure ($y$) because of their position and sharpness.
The temperature dependent magnetic transitions are attributed to the spin state transition of Co$^{3+}$, which is the main focus of the following discussion.
Second, the sudden increase and saturation of the magnetization below 10 T \cite{Hejtmanek2013} may be attributed to the Curie paramagnetism of Pr$^{4+}$ or Co$^{4+}$. 
The third component is the magnetization which is proportional to the magnetic fields, showing little dependence on temperature. 
This component is considered to have other origins such as the Van Vleck susceptibility of Pr$^{4+}$ and Pr$^{3+}$ because they are insensitive to temperature.

The heating effect during the magnetization process may not be as significant as it alters the understanding of the current data.
For example, irreversible heating is expected in magnetization curves in Figs. \ref{fig2}(a)-\ref{fig2}(c) due to the hysteresis loss \cite{Nomura}.
The largest hysteresis loss is expected in the data at $y=0.1$ and $T=5$ K in Fig. \ref{fig2}(c).
In this magnetization curve, we clearly observe the sharp first-order transition in both ascending and descending curves, which is not disturbed by the heating effect.
On the other hand, for cases with $y=0.075$ and 0.0625 at the lowest temperatures, the transitions in the descending curves are blurred compared with the ones in the ascending curves, although irreversible heating is expected to be smaller than that of $y=0.1$
This observation indicates that the smearing effects in the descending curves observed for $y=0.075$ and 0.0625 are not mainly due to the heating effects.
We suspect dynamical effects leading to the sweep rate dependence are more relevant in the present case \cite{Kimura2005}. 

\begin{figure}[b]
\begin{center}
\includegraphics[angle=0, scale=0.6, clip]{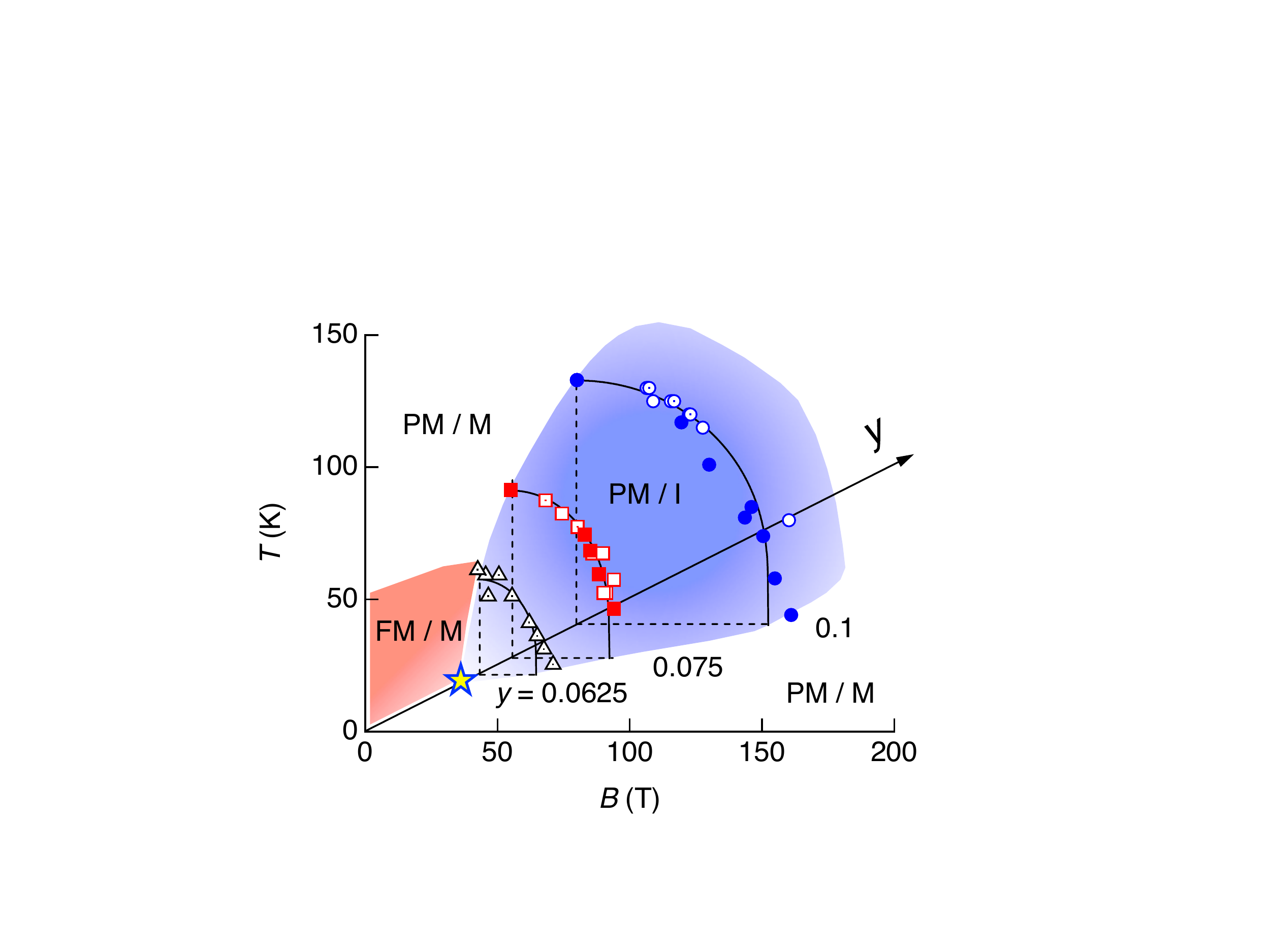} 
\caption{Mean transition fields of (Pr$_{1-y}$Y$_{y}$)$_{0.7}$Ca$_{0.3}$CoO$_{3}$ obtained as a function of temperature and the amount of Y doping $y$. On the $T$-$y$ plane at $B=0$ T, the $T$-$y$ phase diagram for $0<y<0.15$ is depicted based on Refs. [\onlinecite{Tsubouchi2002, Fujita, Naito2010}]. Blue circles, red squares and black triangles indicate transition fields for $y=0.0625$, 0.075 and 0.1, respectively. Open symbols with a center dot indicate that they are obtained using non-destructive pulses. Open symbols without a dot indicate that they are obtained with resistivity measurements using the single-turn coil method adopted from Ref. \cite{Suyeon}. Solid symbols indicate that they are obtained with inductive magnetization measurements using the single-turn coil method. Solid curves are the fittings to the data with the spin-crossover model.}\label{pd}
\end{center}
\end{figure}

We here define the mean transition fields as the average values of the magnetic transition fields in the ascending and descending fields in each magnetization curve in Fig. \ref{fig2}, which are calculated and plotted in the $B$-$T$-$y$ space in Fig. \ref{pd}.
The magnetic transition fields well coincide well with the transition fields of the metal-insulator transitions that were previously reported \cite{Suyeon} as indicated by the open symbols in Fig. \ref{pd}, suggesting strong coupling of the magnetism and conductivity in these systems up to $\sim100$ T.
On the $T$-$y$ plane at $B=0$ T, in Fig. \ref{pd}, the $T$-$y$ phase diagram of (Pr$_{1-y}$Y$_{y}$)$_{0.7}$Ca$_{0.3}$CoO$_{3}$ is depicted according to the previous reports for $0<y<0.15$ \cite{Tsubouchi2004, Naito2010}.
At each $y$ on the respective $B$-$T$ plane, the ground state of the paramagnetic insular phase (PM/I) forms a dome-like region, which is destroyed by either temperature or magnetic field. 
The region outside the nonmagnetic insulator phase is the paramagnetic metallic phase (PM/M), characterized by an increase in the magnetization and conductivity \cite{Suyeon}.
We specifically term the area colored in red in Fig. \ref{pd}  the ferromagnetic metallic phase (FM/M) with  spontaneous magnetization \cite{Tsubouchi2004, Naito2010}.

\section{Discussion}

\begin{figure}
\begin{center}
\includegraphics[angle=0, scale=0.7, clip]{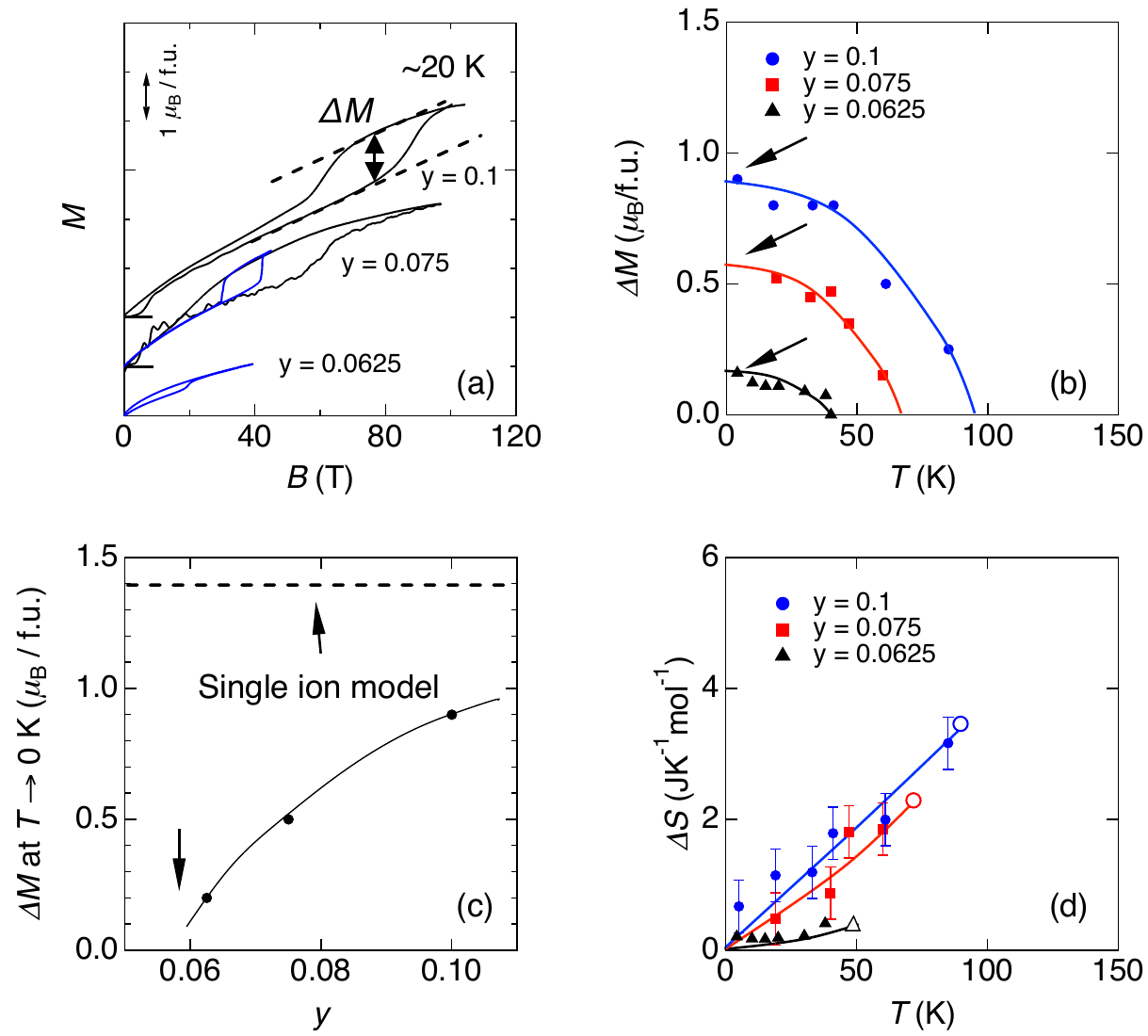} 
\caption{ (a) Magnetization curves of (Pr$_{1-y}$Y$_{y}$)$_{0.7}$Ca$_{0.3}$CoO$_{3}$ at $\sim20$ K selected from Figs. \ref{fig1}. (b) The magnetization increase  at the magnetic transition as a function of temperature. (c) The extrapolated magnetization increase at $T=0$ K deduced in (b).  (d) Increase of entropy $\Delta S$ at the magnetic transition deduced based on the Clausius-Clapeyron relation. The open symbols are reported data measured with heat-capacity measurements at $B=0$ T as summarized in Table \ref{tab}. Solid curves in (b), (c), and (d) are guides for eyes. \label{fig5}}
\end{center}
\end{figure}

In the following discussion, we concentrate on the temperature dependent features of the magnetization curves of (Pr$_{1-y}$Y$_{y}$)$_{0.7}$Ca$_{0.3}$CoO$_{3}$ for $y=0.0625$, 0.075, 0.1, namely, the transition fields $B_{\mathrm{C}}$ (see Fig. \ref{pd}), the magnetization change $\Delta M$ [see Fig. \ref{fig2} and Figs. \ref{fig5}(a) to \ref{fig5}(c)] that are accompanied by the metal-insulator transition.
First, we tentatively analyze  the obtained $B$-$T$-$y$ phase diagram (Fig. \ref{pd}) and the magnetization curves with the spin crossover model of a single ion based on Biernacki and Clerjaud's formulation \cite{Biernacki2005}, which is shown to give a qualitative agreement with the experimental results. 
However, the experimentally obtained $y$ dependence of $\Delta M$ does not agree with the model calculation especially when approaching low temperatures.
It is discussed that the discrepancy may arise from the itinerant magnetism of the paramagnetic metallic phase. 
Further, it is noted that a critical point may be present at low temperature and zero magnetic field with $y$ slightly smaller than 0.0625.
Last, we note the apparent contrast of the present phase diagram to that of LaCoO$_{3}$ and discuss the difference of microscopic origin of those ground states.

\begin{figure}
\begin{center}
\includegraphics[angle=0, scale=0.7, clip]{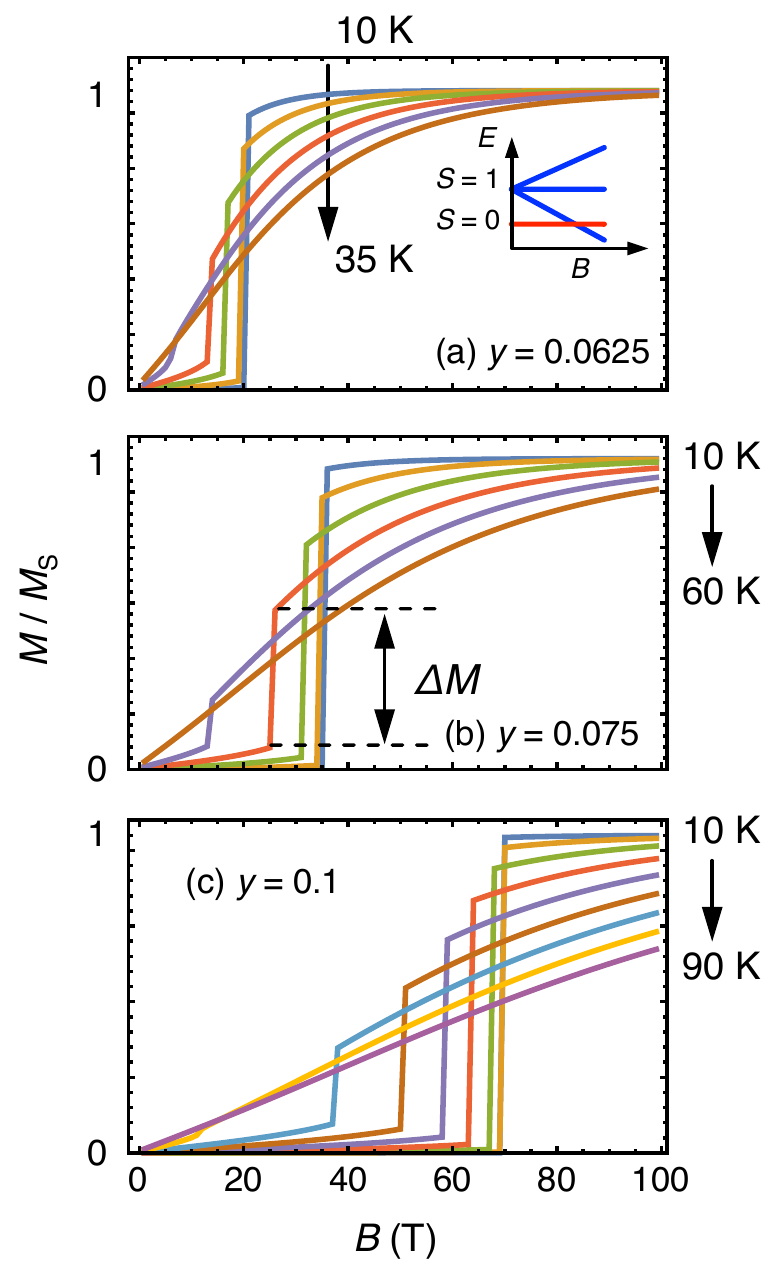} 
\caption{Simulated magnetization curves based on the spin-crossover model of a single ion based on Biernacki and Clejaud's formulation \cite{Biernacki2005}. \label{sim}}
\end{center}
\end{figure}

The spin crossover model used here is based on Biernacki and Clerjaud's formulation \cite{Biernacki2005} where the Boltzmann distribution on a level scheme of a single-ion spin-gap is considered with further modification by mean fields of inter-ion interactions (see the Appendix for details).  The calculated $B_{\mathrm{C}}$ as a function of $T$ using Eq. (\ref{tc1}) (solid curves in Fig. \ref{pd}) are well fitted to the experimentally obtained transition fields assuming the effective crystal field splitting $\Delta'$ of 40, 70, 102 K with $g$-factors of $2.7, 2.8, 2.1$ for $y=0.0625$, 0.075, 0.1, respectively.
The increase in $\Delta'$ with increasing $y$ is understood as the increase of crystal field splitting with increasing chemical pressure. 
The simulated $M$ shown in Fig. \ref{sim} calculated using $M=-\partial F/\partial B$ with Eq. (\ref{eqn}) are in qualitative agreement with the experimental results (Fig. \ref{fig2}), where $B_{\mathrm{C}}$ and $\Delta M$ increase with decreasing temperature or increasing $y$ (chemical pressure corresponding here to the increasing $\Delta'$). 
However, when approaching low temperatures, striking discrepancies in $M$ between the model calculation and the experimental observation are seen.
In the calculated results, $\Delta M$ always converges to the saturation magnetization as shown in Fig. \ref{sim} and summarized in Fig. \ref{fig5}(b).
This is because the temperature dependence of $\Delta M$ in this model originates in the thermal smearing, which become negligible at low temperatures.
In contrast, the experimentally obtained $\Delta M$ converge to values at low temperatures and decrease with decreasing $y$, as seen in Fig. \ref{fig2} and summarized in Fig. \ref{fig5}(a)$-$\ref{fig5}(c). In Fig. \ref{fig5}(c), $\Delta M$ of 1.4$\mu_{\mathrm{B}}$/f.u. at temperature close to 0 K is estimated based on the single ion model where the spin state of 0.7Co$^{3+}$ changes from LS ($S=0$) to IS ($S=1$) with a $g$-factor of 2. The estimated $\Delta M$ is independent of $y$ which highlights well the deviation from the experimentally obtained $\Delta M$, which is largely $y$ dependent.

Possible origins of the $y$ dependence of $\Delta M$ are  (i) a valence change of the Pr ion \cite{Hejtmanek2013} (ii) the inhomogeneous spin-glass behavior of IS Co$^{3+}$, as seen in La$_{1-x}$Sr$_{x}$CoO$_{3}$ with $0<x\leqq0.18$ \cite{Itoh} (iii) itinerant magnetism of Co in analogy to the cases of La$_{1-x}$Sr$_{x}$MnO$_{3}$ with $0.2\leqq x\leqq0.4$ \cite{Saitoh2} and La$_{1-x}$Sr$_{x}$CoO$_{3}$ with $x>0.3$\cite{Saitoh1, Okamoto}.

First, the valence change of Pr ions during the phase transition is discussed as a possible origin of the $y$ dependence of $\Delta M$.
We find below that the $y$ dependence of the valence change is too small to account for the $y$ dependence of the magnetization change in the region of $y$ from 0.0625 to 0.1.
In the crystal field of (Pr$_{1-y}$Y$_{y}$)$_{0.7}$Ca$_{0.3}$CoO$_{3}$, it is estimated that the $^{3}H_{4}$ multiplet of Pr$^{3+}$ (4$f^{2}$) is split into nine nonmagnetic singlets and that the magnetism of the ground-state is represented by an averaged Van Vleck susceptibility of $\chi=0.018$ emu mol$^{-1}$ Oe$^{-1}$ \cite{Novak}.
As for Pr$^{4+}$ (4$f^{1}$), the $^{2}F_{5/2}$ multiplet splits into three energetically well separated Kramers doublets ($> 32$ meV), whose ground doublet is characterized by anisotropic $g$ factor of $g_{x}=3.757$, $g_{y}=0.935$, $g_{z}=0.606$ \cite{Jirak}.
Considering that the ground states are singlet and doublet for Pr$^{3+}$ and Pr$^{4+}$, respectively, the magnetic field will stabilize Pr$^{4+}$ rather than Pr$^{3+}$, inducing the valence change in the direction of Pr$^{3+}$ to Pr$^{4+}$.

On the other hand, if the crystal field effect is overwhelmed by the Zeeman effect at sufficiently high magnetic fields, it is possible that the moment of Pr$^{3+}$ becomes larger than that of Pr$^{4+}$ due to the alternation of the ground state.
In fact, there are excited singlets at 4.5, 14.5, and 16.0 meV above the ground singlet in Pr$^{3+}$ (4$f^{2}$) \cite{Novak} that will be transformed into magnetic eigenstates under ultrahigh magnetic fields.
To roughly estimate the magnetic moments of Pr$^{3+}$ and Pr$^{4+}$ at high magnetic fields, we compare the magnetic moments of free ions.
In this case, the $^{3}H_{4}$ multiplet of Pr$^{3+}$ ($g_{J}=4/5$) has a larger moment of 16/5$\mu_{\mathrm{B}}$ than the $^{2}F_{5/2}$ ($g_{J}=6/7$) multiplet of Pr$^{4+}$ which has a moment of 15/7$\mu_{\mathrm{B}}$.
It is then possible that the valence change in the direction of Pr$^{4+}$ to Pr$^{3+}$ drives the field-induced transition.
If this is the case, the observed $y$ dependence may be explained as the amount of Pr ions that undergoes the valence transition from 4+ to 3+.
In case of the thermally induced phase transition in (Pr$_{1-y}$Y$_{y}$)$_{0.7}$Ca$_{0.3}$CoO$_{3}$, the variation of the amount of the valence change in this $y$ range is small as $\Delta v=$ 0.17, 0.19, 0.2  for $y=0.0625$, 0.075, 0.1, respectively \cite{Hejtmanek2010, Hejtmanek2013}.
Therefore, it is difficult to ascribe the observed $y$ dependent $\Delta M$ directly to the valence transition of Pr ions from 4+ to 3+. 

Second, we discuss that inhomogeneous spin glass behavior of Co moments is unlikely to be the origin of the observed $y$ dependent magnetization change at low temperatures.
In La$_{1-x}$Sr$_{x}$CoO$_{3}$ at $x<0.18$, the inhomogeneous spin glass behavior is observed  \cite{Yamaguchi1995, Raccha}.
If the solid is inhomogeneously magnetized in the field induced spin state transition, the value of $\Delta M$ may be proportional to the amount of the Co ions that undergoes the spin state transition.
In the spin glass regime, the system remains insulating with the percolation limit of $x=0.18$ \cite{Yamaguchi1996}. If $x$ is greater than 0.18, the ferromagnetic drop forms network and system becomes a homogeneous metal \cite{Phelan2014}. 
In the present case of (Pr$_{1-y}$Y$_{y}$)$_{0.7}$Ca$_{0.3}$CoO$_{3}$, all magnetic transitions are accompanied by the metal-insulator transition \cite{Suyeon} even at $y=0.0625$ where $\Delta M$ is only 0.2 $\mu_{\mathrm{B}}$/f.u., \cite{Naito2014, Marysko}, indicating the system is uniformly metallic at above $B_{\mathrm{C}}$ or $T_{\mathrm{C}}$.
Therefore, the idea of inhomogeneous occupation of the magnetic spin states in (Pr$_{1-y}$Y$_{y}$)$_{0.7}$Ca$_{0.3}$CoO$_{3}$ in the metallic phase is not well supported.

Third, we consider the possibility where the magnetization of the paramagnetic metallic phase is determined by the itinerant magnetism of the Co$^{3+}$ band.
As noted in Ref. \cite{Knizek2010}, the phase transition of (Pr$_{1-y}$Y$_{y}$)$_{0.7}$Ca$_{0.3}$CoO$_{3}$ with decreasing temperature is analogues to the virtual transition from the ferromagnetic metal La$_{1-x}$Sr$_{x}$CoO$_{3}$ to the diamagnetic insulator LaCoO$_{3}$.
The double exchange mechanism with Co$^{3+}$ in IS state is considered in play to form the ferromagnetic metallic phase both in La$_{1-x}$Sr$_{x}$CoO$_{3}$ with $x>0.3$ and La$_{1-x}$Sr$_{x}$MnO$_{3}$ with $0.2\leqq x\leqq0.4$ \cite{Saitoh1, Saitoh2, Okamoto}.
The ferromagnetism in La$_{1-x}$Sr$_{x}$CoO$_{3}$ is discussed to be characteristic as compared to the typical double exchange system of La$_{1-x}$Sr$_{x}$MnO$_{3}$.
The ferromagnetic moment of La$_{1-x}$Sr$_{x}$CoO$_{3}$ is smaller than its saturation moment whereas in La$_{1-x}$Sr$_{x}$MnO$_{3}$ the saturation moment is achieved.
It was discussed that this difference arise from the fact that La$_{1-x}$Sr$_{x}$CoO$_{3}$ has larger $d$-$p$ hybridization with better itineracy evidenced by sharper Fermi edge than the case of La$_{1-x}$Sr$_{x}$MnO$_{3}$ \cite{Saitoh2} and that La$_{1-x}$Sr$_{x}$CoO$_{3}$ is not a typical ferromagnet with double exchange mechanism but a more itinerant ferromagnet \cite{Saitoh1, Okamoto}.

\begin{table*}
\caption{Amount of Y doping ($y$), Transition temperature $T_{\mathrm{C}}$ at $B=0$ T, Change of entropy $\Delta S$at $B=0 \mathrm{\ T}$, Deduced enthalpy change $\Delta H_{B=0 \mathrm{\ T}}=T_{\mathrm{C}} \Delta S$, Transition field $B_{\mathrm{C}}$ at $T\rightarrow0 \mathrm{\ K}$, Change of magnetization $\Delta M$ at $T\rightarrow0 \mathrm{\ K}$, Deduced enthalpy change $\Delta H_{T\rightarrow0 \mathrm{\ K}}=B_{\mathrm{C}} \Delta M$, Crystal field gap $\Delta'=T_{\mathrm{C}} \ln 3$ used in the spin crossover model}
\label{tab}
\begin{ruledtabular}
\begin{tabular}{llllllll}

$y$&$T_{\mathrm{C}}$ [K]&$\Delta S$ [J/K mol]&$\Delta H_{B=0 \mathrm{\ T}}$  [K/f.u.]&$B_{\mathrm{C}}$ [T]& $\Delta M$  [$\mu_{\mathrm{B}}/\mathrm{f.u.}$] & $\Delta H_{T\rightarrow0 \mathrm{\ K}}$   [K/f.u.] & $\Delta'$ [K/f.u.]\\

 \hline

0.0625 & 40\footnote{Ref. \cite{Hejtmanek2013}} & 0.28\footnote{Ref. \cite{Naito2014}} & 1.35 &17 &  0.15 & 2.55 & 44 \\
0.075 & 64$^{\text{a}}$ & 2.17\footnote{Ref. \cite{Hejtmanek2010}} & 16.7 & 40 & 0.51 & 20.4 & 70.3 \\
0.1 & 93$^{\text{a}}$ & $>3.6$\footnote{Ref. \cite{HejtmanekPrivate}} & 41.6 & 80 & 0.9 & 72 & 102 \\
0.15 & 132$^{\text{a}}$ & 4.78$^{\text{c}}$ & 75.9 & - & - & -  \\

\end{tabular}
\end{ruledtabular}
\end{table*}

In the present case of  (Pr$_{1-y}$Y$_{y}$)$_{0.7}$Ca$_{0.3}$CoO$_{3}$, we observed increasing $\Delta M$ with increasing chemical pressure, $y$ [Figs. \ref{fig5}(a), \ref{fig5}(b), \ref{fig5}(c)].
For (Pr$_{1-y}$Y$_{y}$)$_{0.7}$Ca$_{0.3}$CoO$_{3}$, the pressure mainly affects the bond length and do not alter the bond angles of Co-O-Co \cite{Fujita}.
With increasing pressure, therefore, the strength of $d$-$p$ hybridization is expected to increases, which results in transforming the system from being similar to La$_{1-x}$Sr$_{x}$MnO$_{3}$ to being similar to La$_{1-x}$Sr$_{x}$CoO$_{3}$.
This will decrease $\Delta M$ with increasing $y$, being in disagreement with the observation. 
We note that the observed large chemical pressure dependence of the magnetization in the paramagnetic metal phase is not similar to the previous double exchange systems where less significant pressure effects are observed \cite{Morimoto1995, Morimoto1997, Lengsdorf, Fita}. 

Rather, it appears more plausible to consider the Stoner model, where exchange splitting of the band is determined by the strength of the exchange interaction.
In the present case, the origin of the exchange interaction is the double exchange mechanism $J\propto t_{0}\cos(\theta/2)$, where $t_{0}$ and $\theta$ are transfer integral and relative angle of spins at adjacent sites.
By applying pressure, $t_{0}$ is increased by stronger hybridization, resulting in the larger exchange interactions.
This will leads to the enhanced exchange splitting of the spin bands.
This idea is consistent with the observed $y$ dependence of $\Delta M$.

Here, we note the possible existence of the critical point at low temperatures in Fig. \ref{pd}.
One may notice that $\Delta M$ at $T\rightarrow0$ K becomes negligibly small with decreasing $y$ as seen in Fig. \ref{fig5}(c), indicating that the second order phase transition is realized at $y$ slightly smaller than 0.0625.
This trend is also the case for the values of $\Delta S$ at $B=0$ T reported by heat capacity measurements as shown in Table \ref{tab}.
It is therefore worth noting the possible existence of the critical point at around the star mark in Fig. \ref{pd}.
The possible critical point separates the ferromagnetic metallic phase and the paramagnetic insulating phase.
It is located at very low temperature.
Therefore, it may be important to consider the ferromagnetic fluctuation, the fluctuation between itineracy and localization, and the emerging quantum criticality at the adjacent region.
We note that the possible critical point can be explored by varying the applied pressure on the samples with $y<0.0625$.

We further estimated the chemical pressure ($y$) dependence of the latent heat at the temperature induced phase transitions, which is defined as the enthalpy change $\Delta H_{B=0 \mathrm{\ T}}$. And $\Delta H_{T\rightarrow0 \mathrm{\ K}}$ at the magnetic field induce phase transitions is also calculated. Enthalpy of the magnetic system is defined as 
\begin{equation}
\Delta H = T\Delta S + B\Delta M.
\label{enthalpy}
\end{equation}
Eq. (\ref{enthalpy}) is reduced to $\Delta H_{B=0 \mathrm{\ T}} = T_{\mathrm{C}}\Delta S$ and $\Delta H_{T\rightarrow0 \mathrm{\ K}} = B_{\mathrm{C}}\Delta M$ at respective conditions.
We estimated each values of $\Delta H$ based on the reported values of $T_{\mathrm{C}}$ and $\Delta S$ at $B=0$ T, and the obtained values of $\Delta M$ and $B_{\mathrm{C}}$ at $T\rightarrow0$ K, respectively. 
The results are summarized in Table \ref{tab}.
The vanishingly decreasing latent heat with decreasing $y$ directly indicates the change from the first order phase transition to the second order phase transition.
The value of $\Delta H_{T\rightarrow0 \mathrm{\ K}}$ also becomes negligibly small, being in good agreement with the above result. This trend is not reproduced in the estimated crystal field gap $\Delta '$ in the spin crossover model.

Lastly we note that the obtained phase diagram has the dome like structure at around its ground state, indicating that the low entropy phase is at its ground state (Fig. \ref{pd}).
The entropy change at the field-induced phase transitions $\Delta S$ are also quantitatively deduced  as shown in Fig. \ref{fig5}(d) based on Clausius-Clapeyron relation, $dB/dT=-\Delta S/\Delta M$, and the experimentally obtained $dB/dT$ in Fig. \ref{pd} and $\Delta M$ in Fig. \ref{fig5}(b). 
Whereas, in LaCoO$_{3}$ the high magnetic field induced phase has phase boundary of positive slope at $B>100$ T indicating that the low entropy ordered phase is at the high magnetic field induced phase \cite{Ikeda}, which is strikingly contrary to the present case.
The obtained phase diagram in Fig. \ref{pd} and the entropy change $\Delta S$ for (Pr$_{1-y}$Y$_{y}$)$_{0.7}$Ca$_{0.3}$CoO$_{3}$ in the present study shows quite obvious contrasts to those obtained for LaCoO$_{3}$ in the previous study \cite{Ikeda}. 
The difference between the phase diagrams of (Pr$_{1-y}$Y$_{y}$)$_{0.7}$Ca$_{0.3}$CoO$_{3}$ and LaCoO$_{3}$ may suggest the difference of microscopic nature of the phase transition.

Recently, excitonic condensation is proposed to be realized in the systems with the spin state degree of freedom to account for the unusual insulating phase in Pr based perovskite cobalt family and LaCoO$_{3}$ \cite{Kunes1, Kaneko, Nasu}.
More recently, Sotnikov and Kune\v{s}\cite{Sotnikov}  and Tatsuno \textit{et al.} \cite{Tatsuno} independently claim the possibility of the magnetic field induced excitonic condensation \cite{Sotnikov, Tatsuno} to account for the recently found high-field phases of LaCoO$_{3}$ \cite{Ikeda}.
Tatsuno \textit{et al.} claim that in case of LaCoO$_{3}$ the field-induced phase transition takes place  in succession as from LS $\rightarrow$ excitonic condensation (EC)  $\rightarrow$ LS/HS  $\rightarrow$  EC  $\rightarrow$  HS phases \cite{Tatsuno}.
Then, one fascinating and united picture describing the $B$-$T$ phase diagrams of LaCoO$_{3}$ and (Pr$_{1-y}$Y$_{y}$)$_{0.7}$Ca$_{0.3}$CoO$_{3}$ may be that LaCoO$_{3}$ is LS at its ground state whereas (Pr$_{1-y}$Y$_{y}$)$_{0.7}$Ca$_{0.3}$CoO$_{3}$ is in the EC phase at its ground state. 
Theoretical suggestion that both LS and EC can be destroyed by the magnetic field may explain the field effects for both materials.
Further experimental verification is needed to test those theories.

\section{Conclusion}

We have carried out the magnetization measurements of (Pr$_{1-y}$Y$_{y}$)$_{0.7}$Ca$_{0.3}$CoO$_{3}$ with $y=0.0625$, 0.075, 0.1 at high magnetic fields up to 140 T and observed the magnetic phase transition. The $B$-$T$ phase diagram obtained from the temperature dependence and the chemical pressure dependence of the transition fields are in good agreement with the single ion model experiencing the spin crossover, where the ground state is the low entropy phase. This is in contrast with the case of the previously investigated LaCoO$_{3}$ \cite{Ikeda} where the low entropy ordered phases emerges at high magnetic fields. On the other hand, the qualitative discrepancy of the magnetization jump $\Delta M$ as a function of $y$ in experiment and the single ion model has been found. The origin of the discrepancy is attributed to the itinerant magnetism of Co ion. Further, the existence of the critical point at low temperature is indicated by the vanishingly small $\Delta M$, $\Delta S$ and latent heat for decreasing $y$ at low temperatures.

\section{Akwknowledgement}
The authors acknowledge J. Nasu, S. Ishihara for fruitful discussions, T. Nomura and Y. Kohama for useful comments, and K. Kindo and S. Takeyama for experimental supports. This work was supported by JSPS KAKENHI Grant-in-Aid for Young Scientists (B) Grant No. 16K17738.

\bibliography{pycco}

\begin{thebibliography}{44}%
\makeatletter
\providecommand \@ifxundefined [1]{%
 \@ifx{#1\undefined}
}%
\providecommand \@ifnum [1]{%
 \ifnum #1\expandafter \@firstoftwo
 \else \expandafter \@secondoftwo
 \fi
}%
\providecommand \@ifx [1]{%
 \ifx #1\expandafter \@firstoftwo
 \else \expandafter \@secondoftwo
 \fi
}%
\providecommand \natexlab [1]{#1}%
\providecommand \enquote  [1]{``#1''}%
\providecommand \bibnamefont  [1]{#1}%
\providecommand \bibfnamefont [1]{#1}%
\providecommand \citenamefont [1]{#1}%
\providecommand \href@noop [0]{\@secondoftwo}%
\providecommand \href [0]{\begingroup \@sanitize@url \@href}%
\providecommand \@href[1]{\@@startlink{#1}\@@href}%
\providecommand \@@href[1]{\endgroup#1\@@endlink}%
\providecommand \@sanitize@url [0]{\catcode `\\12\catcode `\$12\catcode
  `\&12\catcode `\#12\catcode `\^12\catcode `\_12\catcode `\%12\relax}%
\providecommand \@@startlink[1]{}%
\providecommand \@@endlink[0]{}%
\providecommand \url  [0]{\begingroup\@sanitize@url \@url }%
\providecommand \@url [1]{\endgroup\@href {#1}{\urlprefix }}%
\providecommand \urlprefix  [0]{URL }%
\providecommand \Eprint [0]{\href }%
\providecommand \doibase [0]{http://dx.doi.org/}%
\providecommand \selectlanguage [0]{\@gobble}%
\providecommand \bibinfo  [0]{\@secondoftwo}%
\providecommand \bibfield  [0]{\@secondoftwo}%
\providecommand \translation [1]{[#1]}%
\providecommand \BibitemOpen [0]{}%
\providecommand \bibitemStop [0]{}%
\providecommand \bibitemNoStop [0]{.\EOS\space}%
\providecommand \EOS [0]{\spacefactor3000\relax}%
\providecommand \BibitemShut  [1]{\csname bibitem#1\endcsname}%
\let\auto@bib@innerbib\@empty
%</preamble>
\bibitem [{\citenamefont {Asai}\ \emph {et~al.}(1994)\citenamefont {Asai},
  \citenamefont {Yokokura}, \citenamefont {Nishimori}, \citenamefont {Chou},
  \citenamefont {Tranquada}, \citenamefont {Shirane}, \citenamefont {Higuchi},
  \citenamefont {Okajima},\ and\ \citenamefont {Kohn}}]{Asai1994}%
  \BibitemOpen
  \bibfield  {author} {\bibinfo {author} {\bibfnamefont {K.}~\bibnamefont
  {Asai}}, \bibinfo {author} {\bibfnamefont {O.}~\bibnamefont {Yokokura}},
  \bibinfo {author} {\bibfnamefont {N.}~\bibnamefont {Nishimori}}, \bibinfo
  {author} {\bibfnamefont {H.}~\bibnamefont {Chou}}, \bibinfo {author}
  {\bibfnamefont {J.~M.}\ \bibnamefont {Tranquada}}, \bibinfo {author}
  {\bibfnamefont {G.}~\bibnamefont {Shirane}}, \bibinfo {author} {\bibfnamefont
  {S.}~\bibnamefont {Higuchi}}, \bibinfo {author} {\bibfnamefont
  {Y.}~\bibnamefont {Okajima}}, \ and\ \bibinfo {author} {\bibfnamefont
  {K.}~\bibnamefont {Kohn}},\ }\href {\doibase 10.1103/PhysRevB.50.3025}
  {\bibfield  {journal} {\bibinfo  {journal} {Phys. Rev. B}\ }\textbf {\bibinfo
  {volume} {50}},\ \bibinfo {pages} {3025} (\bibinfo {year}
  {1994})}\BibitemShut {NoStop}%
\bibitem [{\citenamefont {Saitoh}\ \emph
  {et~al.}(1997{\natexlab{a}})\citenamefont {Saitoh}, \citenamefont {Mizokawa},
  \citenamefont {Fujimori}, \citenamefont {Abbate}, \citenamefont {Takeda},\
  and\ \citenamefont {Takano}}]{Saitoh}%
  \BibitemOpen
  \bibfield  {author} {\bibinfo {author} {\bibfnamefont {T.}~\bibnamefont
  {Saitoh}}, \bibinfo {author} {\bibfnamefont {T.}~\bibnamefont {Mizokawa}},
  \bibinfo {author} {\bibfnamefont {A.}~\bibnamefont {Fujimori}}, \bibinfo
  {author} {\bibfnamefont {M.}~\bibnamefont {Abbate}}, \bibinfo {author}
  {\bibfnamefont {Y.}~\bibnamefont {Takeda}}, \ and\ \bibinfo {author}
  {\bibfnamefont {M.}~\bibnamefont {Takano}},\ }\href {\doibase
  10.1103/PhysRevB.55.4257} {\bibfield  {journal} {\bibinfo  {journal} {Phys.
  Rev. B}\ }\textbf {\bibinfo {volume} {55}},\ \bibinfo {pages} {4257}
  (\bibinfo {year} {1997}{\natexlab{a}})}\BibitemShut {NoStop}%
\bibitem [{\citenamefont {Raccah}\ and\ \citenamefont
  {Goodenough}(1967)}]{Raccha}%
  \BibitemOpen
  \bibfield  {author} {\bibinfo {author} {\bibfnamefont {P.~M.}\ \bibnamefont
  {Raccah}}\ and\ \bibinfo {author} {\bibfnamefont {J.~B.}\ \bibnamefont
  {Goodenough}},\ }\href {\doibase 10.1103/PhysRev.155.932} {\bibfield
  {journal} {\bibinfo  {journal} {Phys. Rev.}\ }\textbf {\bibinfo {volume}
  {155}},\ \bibinfo {pages} {932} (\bibinfo {year} {1967})}\BibitemShut
  {NoStop}%
\bibitem [{\citenamefont {Korotin}\ \emph {et~al.}(1996)\citenamefont
  {Korotin}, \citenamefont {Ezhov}, \citenamefont {Solovyev}, \citenamefont
  {Anisimov}, \citenamefont {Khomskii},\ and\ \citenamefont
  {Sawatzky}}]{Korotin}%
  \BibitemOpen
  \bibfield  {author} {\bibinfo {author} {\bibfnamefont {M.~A.}\ \bibnamefont
  {Korotin}}, \bibinfo {author} {\bibfnamefont {S.~Y.}\ \bibnamefont {Ezhov}},
  \bibinfo {author} {\bibfnamefont {I.~V.}\ \bibnamefont {Solovyev}}, \bibinfo
  {author} {\bibfnamefont {V.~I.}\ \bibnamefont {Anisimov}}, \bibinfo {author}
  {\bibfnamefont {D.~I.}\ \bibnamefont {Khomskii}}, \ and\ \bibinfo {author}
  {\bibfnamefont {G.~A.}\ \bibnamefont {Sawatzky}},\ }\href {\doibase
  10.1103/PhysRevB.54.5309} {\bibfield  {journal} {\bibinfo  {journal} {Phys.
  Rev. B}\ }\textbf {\bibinfo {volume} {54}},\ \bibinfo {pages} {5309}
  (\bibinfo {year} {1996})}\BibitemShut {NoStop}%
\bibitem [{\citenamefont {Tsubouchi}\ \emph {et~al.}(2002)\citenamefont
  {Tsubouchi}, \citenamefont {Ky\^omen}, \citenamefont {Itoh}, \citenamefont
  {Ganguly}, \citenamefont {Oguni}, \citenamefont {Shimojo}, \citenamefont
  {Morii},\ and\ \citenamefont {Ishii}}]{Tsubouchi2002}%
  \BibitemOpen
  \bibfield  {author} {\bibinfo {author} {\bibfnamefont {S.}~\bibnamefont
  {Tsubouchi}}, \bibinfo {author} {\bibfnamefont {T.}~\bibnamefont {Ky\^omen}},
  \bibinfo {author} {\bibfnamefont {M.}~\bibnamefont {Itoh}}, \bibinfo {author}
  {\bibfnamefont {P.}~\bibnamefont {Ganguly}}, \bibinfo {author} {\bibfnamefont
  {M.}~\bibnamefont {Oguni}}, \bibinfo {author} {\bibfnamefont
  {Y.}~\bibnamefont {Shimojo}}, \bibinfo {author} {\bibfnamefont
  {Y.}~\bibnamefont {Morii}}, \ and\ \bibinfo {author} {\bibfnamefont
  {Y.}~\bibnamefont {Ishii}},\ }\href {\doibase 10.1103/PhysRevB.66.052418}
  {\bibfield  {journal} {\bibinfo  {journal} {Phys. Rev. B}\ }\textbf {\bibinfo
  {volume} {66}},\ \bibinfo {pages} {052418} (\bibinfo {year}
  {2002})}\BibitemShut {NoStop}%
\bibitem [{\citenamefont {Kn\'{i}\v{z}ek}\ \emph {et~al.}(2010)\citenamefont
  {Kn\'{i}\v{z}ek}, \citenamefont {Hejtm\'{a}nek}, \citenamefont {Nov\'{a}k},\
  and\ \citenamefont {Jir\'{a}k}}]{Knizek2010}%
  \BibitemOpen
  \bibfield  {author} {\bibinfo {author} {\bibfnamefont {K.}~\bibnamefont
  {Kn\'{i}\v{z}ek}}, \bibinfo {author} {\bibfnamefont {J.}~\bibnamefont
  {Hejtm\'{a}nek}}, \bibinfo {author} {\bibfnamefont {P.}~\bibnamefont
  {Nov\'{a}k}}, \ and\ \bibinfo {author} {\bibfnamefont {Z.}~\bibnamefont
  {Jir\'{a}k}},\ }\href {\doibase 10.1103/PhysRevB.81.155113} {\bibfield
  {journal} {\bibinfo  {journal} {Phys. Rev. B}\ }\textbf {\bibinfo {volume}
  {81}},\ \bibinfo {pages} {155113} (\bibinfo {year} {2010})}\BibitemShut
  {NoStop}%
\bibitem [{\citenamefont {Saitoh}\ \emph
  {et~al.}(1997{\natexlab{b}})\citenamefont {Saitoh}, \citenamefont {Mizokawa},
  \citenamefont {Fujimori}, \citenamefont {Abbate}, \citenamefont {Takeda},\
  and\ \citenamefont {Takano}}]{Saitoh1}%
  \BibitemOpen
  \bibfield  {author} {\bibinfo {author} {\bibfnamefont {T.}~\bibnamefont
  {Saitoh}}, \bibinfo {author} {\bibfnamefont {T.}~\bibnamefont {Mizokawa}},
  \bibinfo {author} {\bibfnamefont {A.}~\bibnamefont {Fujimori}}, \bibinfo
  {author} {\bibfnamefont {M.}~\bibnamefont {Abbate}}, \bibinfo {author}
  {\bibfnamefont {Y.}~\bibnamefont {Takeda}}, \ and\ \bibinfo {author}
  {\bibfnamefont {M.}~\bibnamefont {Takano}},\ }\href {\doibase
  10.1103/PhysRevB.56.1290} {\bibfield  {journal} {\bibinfo  {journal} {Phys.
  Rev. B}\ }\textbf {\bibinfo {volume} {56}},\ \bibinfo {pages} {1290}
  (\bibinfo {year} {1997}{\natexlab{b}})}\BibitemShut {NoStop}%
\bibitem [{\citenamefont {Fujita}\ \emph {et~al.}(2004)\citenamefont {Fujita},
  \citenamefont {Miyashita}, \citenamefont {Yasui}, \citenamefont {Kobayashi},
  \citenamefont {Sato}, \citenamefont {Nishibori}, \citenamefont {Sakata},
  \citenamefont {Shimojo}, \citenamefont {Igawa}, \citenamefont {Ishii},
  \citenamefont {Kakurai}, \citenamefont {Adachi}, \citenamefont {Ohishi},\
  and\ \citenamefont {Takata}}]{Fujita}%
  \BibitemOpen
  \bibfield  {author} {\bibinfo {author} {\bibfnamefont {T.}~\bibnamefont
  {Fujita}}, \bibinfo {author} {\bibfnamefont {T.}~\bibnamefont {Miyashita}},
  \bibinfo {author} {\bibfnamefont {Y.}~\bibnamefont {Yasui}}, \bibinfo
  {author} {\bibfnamefont {Y.}~\bibnamefont {Kobayashi}}, \bibinfo {author}
  {\bibfnamefont {M.}~\bibnamefont {Sato}}, \bibinfo {author} {\bibfnamefont
  {E.}~\bibnamefont {Nishibori}}, \bibinfo {author} {\bibfnamefont
  {M.}~\bibnamefont {Sakata}}, \bibinfo {author} {\bibfnamefont
  {Y.}~\bibnamefont {Shimojo}}, \bibinfo {author} {\bibfnamefont
  {N.}~\bibnamefont {Igawa}}, \bibinfo {author} {\bibfnamefont
  {Y.}~\bibnamefont {Ishii}}, \bibinfo {author} {\bibfnamefont
  {K.}~\bibnamefont {Kakurai}}, \bibinfo {author} {\bibfnamefont
  {T.}~\bibnamefont {Adachi}}, \bibinfo {author} {\bibfnamefont
  {Y.}~\bibnamefont {Ohishi}}, \ and\ \bibinfo {author} {\bibfnamefont
  {M.}~\bibnamefont {Takata}},\ }\href {\doibase 10.1143/JPSJ.73.1987}
  {\bibfield  {journal} {\bibinfo  {journal} {J. Phys. Soc. Jpn.}\ }\textbf
  {\bibinfo {volume} {73}},\ \bibinfo {pages} {1987} (\bibinfo {year}
  {2004})}\BibitemShut {NoStop}%
\bibitem [{\citenamefont {Tsubouchi}\ \emph {et~al.}(2004)\citenamefont
  {Tsubouchi}, \citenamefont {Ky\^omen}, \citenamefont {Itoh},\ and\
  \citenamefont {Oguni}}]{Tsubouchi2004}%
  \BibitemOpen
  \bibfield  {author} {\bibinfo {author} {\bibfnamefont {S.}~\bibnamefont
  {Tsubouchi}}, \bibinfo {author} {\bibfnamefont {T.}~\bibnamefont {Ky\^omen}},
  \bibinfo {author} {\bibfnamefont {M.}~\bibnamefont {Itoh}}, \ and\ \bibinfo
  {author} {\bibfnamefont {M.}~\bibnamefont {Oguni}},\ }\href {\doibase
  10.1103/PhysRevB.69.144406} {\bibfield  {journal} {\bibinfo  {journal} {Phys.
  Rev. B}\ }\textbf {\bibinfo {volume} {69}},\ \bibinfo {pages} {144406}
  (\bibinfo {year} {2004})}\BibitemShut {NoStop}%
\bibitem [{\citenamefont {Naito}\ \emph {et~al.}(2010)\citenamefont {Naito},
  \citenamefont {Sasaki},\ and\ \citenamefont {Fujishiro}}]{Naito2010}%
  \BibitemOpen
  \bibfield  {author} {\bibinfo {author} {\bibfnamefont {T.}~\bibnamefont
  {Naito}}, \bibinfo {author} {\bibfnamefont {H.}~\bibnamefont {Sasaki}}, \
  and\ \bibinfo {author} {\bibfnamefont {H.}~\bibnamefont {Fujishiro}},\ }\href
  {\doibase 10.1143/Jpsj.79.034710} {\bibfield  {journal} {\bibinfo  {journal}
  {J. Phys. Soc. Jpn.}\ }\textbf {\bibinfo {volume} {79}},\ \bibinfo {pages}
  {034710} (\bibinfo {year} {2010})}\BibitemShut {NoStop}%
\bibitem [{\citenamefont {Phelan}\ \emph {et~al.}(2014)\citenamefont {Phelan},
  \citenamefont {Bhatti}, \citenamefont {Taylor}, \citenamefont {Wang},\ and\
  \citenamefont {Leighton}}]{Phelan2014}%
  \BibitemOpen
  \bibfield  {author} {\bibinfo {author} {\bibfnamefont {D.}~\bibnamefont
  {Phelan}}, \bibinfo {author} {\bibfnamefont {K.~P.}\ \bibnamefont {Bhatti}},
  \bibinfo {author} {\bibfnamefont {M.}~\bibnamefont {Taylor}}, \bibinfo
  {author} {\bibfnamefont {S.}~\bibnamefont {Wang}}, \ and\ \bibinfo {author}
  {\bibfnamefont {C.}~\bibnamefont {Leighton}},\ }\href {\doibase
  10.1103/PhysRevB.89.184427} {\bibfield  {journal} {\bibinfo  {journal} {Phys.
  Rev. B}\ }\textbf {\bibinfo {volume} {89}},\ \bibinfo {pages} {184427}
  (\bibinfo {year} {2014})}\BibitemShut {NoStop}%
\bibitem [{\citenamefont {Kune\v{s}}\ and\ \citenamefont
  {Augustinsk\'{y}}(2014)}]{Kunes1}%
  \BibitemOpen
  \bibfield  {author} {\bibinfo {author} {\bibfnamefont {J.}~\bibnamefont
  {Kune\v{s}}}\ and\ \bibinfo {author} {\bibfnamefont {P.}~\bibnamefont
  {Augustinsk\'{y}}},\ }\href {\doibase 10.1103/PhysRevB.90.235112} {\bibfield
  {journal} {\bibinfo  {journal} {Phys. Rev. B}\ }\textbf {\bibinfo {volume}
  {90}},\ \bibinfo {pages} {235112} (\bibinfo {year} {2014})}\BibitemShut
  {NoStop}%
\bibitem [{\citenamefont {Kaneko}\ \emph {et~al.}(2012)\citenamefont {Kaneko},
  \citenamefont {Seki},\ and\ \citenamefont {Ohta}}]{Kaneko}%
  \BibitemOpen
  \bibfield  {author} {\bibinfo {author} {\bibfnamefont {T.}~\bibnamefont
  {Kaneko}}, \bibinfo {author} {\bibfnamefont {K.}~\bibnamefont {Seki}}, \ and\
  \bibinfo {author} {\bibfnamefont {Y.}~\bibnamefont {Ohta}},\ }\href {\doibase
  10.1103/PhysRevB.85.165135} {\bibfield  {journal} {\bibinfo  {journal} {Phys.
  Rev. B}\ }\textbf {\bibinfo {volume} {85}},\ \bibinfo {pages} {165135}
  (\bibinfo {year} {2012})}\BibitemShut {NoStop}%
\bibitem [{\citenamefont {Nasu}\ \emph {et~al.}(2016)\citenamefont {Nasu},
  \citenamefont {Watanabe}, \citenamefont {Naka},\ and\ \citenamefont
  {Ishihara}}]{Nasu}%
  \BibitemOpen
  \bibfield  {author} {\bibinfo {author} {\bibfnamefont {J.}~\bibnamefont
  {Nasu}}, \bibinfo {author} {\bibfnamefont {T.}~\bibnamefont {Watanabe}},
  \bibinfo {author} {\bibfnamefont {M.}~\bibnamefont {Naka}}, \ and\ \bibinfo
  {author} {\bibfnamefont {S.}~\bibnamefont {Ishihara}},\ }\href {\doibase
  10.1103/PhysRevB.93.205136} {\bibfield  {journal} {\bibinfo  {journal} {Phys.
  Rev. B}\ }\textbf {\bibinfo {volume} {93}},\ \bibinfo {pages} {205136}
  (\bibinfo {year} {2016})}\BibitemShut {NoStop}%
\bibitem [{\citenamefont {Sato}\ \emph {et~al.}(2009)\citenamefont {Sato},
  \citenamefont {Matsuo}, \citenamefont {Kindo}, \citenamefont {Kobayashi},\
  and\ \citenamefont {Asai}}]{Sato2009}%
  \BibitemOpen
  \bibfield  {author} {\bibinfo {author} {\bibfnamefont {K.}~\bibnamefont
  {Sato}}, \bibinfo {author} {\bibfnamefont {A.}~\bibnamefont {Matsuo}},
  \bibinfo {author} {\bibfnamefont {K.}~\bibnamefont {Kindo}}, \bibinfo
  {author} {\bibfnamefont {Y.}~\bibnamefont {Kobayashi}}, \ and\ \bibinfo
  {author} {\bibfnamefont {K.}~\bibnamefont {Asai}},\ }\href {\doibase
  10.1143/JPSJ.80.104702} {\bibfield  {journal} {\bibinfo  {journal} {J. Phys.
  Soc. Jpn.}\ }\textbf {\bibinfo {volume} {78}},\ \bibinfo {pages} {093702}
  (\bibinfo {year} {2009})}\BibitemShut {NoStop}%
\bibitem [{\citenamefont {Altarawneh}\ \emph {et~al.}(2012)\citenamefont
  {Altarawneh}, \citenamefont {Chern}, \citenamefont {Harrison}, \citenamefont
  {Batista}, \citenamefont {Uchida}, \citenamefont {Jaime}, \citenamefont
  {Rickel}, \citenamefont {Crooker}, \citenamefont {Mielke}, \citenamefont
  {Betts}, \citenamefont {Mitchell},\ and\ \citenamefont {Hoch}}]{Moaz}%
  \BibitemOpen
  \bibfield  {author} {\bibinfo {author} {\bibfnamefont {M.~M.}\ \bibnamefont
  {Altarawneh}}, \bibinfo {author} {\bibfnamefont {G.~W.}\ \bibnamefont
  {Chern}}, \bibinfo {author} {\bibfnamefont {N.}~\bibnamefont {Harrison}},
  \bibinfo {author} {\bibfnamefont {C.~D.}\ \bibnamefont {Batista}}, \bibinfo
  {author} {\bibfnamefont {A.}~\bibnamefont {Uchida}}, \bibinfo {author}
  {\bibfnamefont {M.}~\bibnamefont {Jaime}}, \bibinfo {author} {\bibfnamefont
  {D.~G.}\ \bibnamefont {Rickel}}, \bibinfo {author} {\bibfnamefont {S.~A.}\
  \bibnamefont {Crooker}}, \bibinfo {author} {\bibfnamefont {C.~H.}\
  \bibnamefont {Mielke}}, \bibinfo {author} {\bibfnamefont {J.~B.}\
  \bibnamefont {Betts}}, \bibinfo {author} {\bibfnamefont {J.~F.}\ \bibnamefont
  {Mitchell}}, \ and\ \bibinfo {author} {\bibfnamefont {M.~J.~R.}\ \bibnamefont
  {Hoch}},\ }\href {\doibase 10.1103/PhysRevLett.109.037201} {\bibfield
  {journal} {\bibinfo  {journal} {Phys. Rev. Lett.}\ }\textbf {\bibinfo
  {volume} {109}},\ \bibinfo {pages} {037201} (\bibinfo {year}
  {2012})}\BibitemShut {NoStop}%
\bibitem [{\citenamefont {Rotter}\ \emph {et~al.}(2014)\citenamefont {Rotter},
  \citenamefont {Wang}, \citenamefont {Boothroyd}, \citenamefont {Prabhakaran},
  \citenamefont {Tanaka},\ and\ \citenamefont {Doerr}}]{Rotter}%
  \BibitemOpen
  \bibfield  {author} {\bibinfo {author} {\bibfnamefont {M.}~\bibnamefont
  {Rotter}}, \bibinfo {author} {\bibfnamefont {Z.~S.}\ \bibnamefont {Wang}},
  \bibinfo {author} {\bibfnamefont {A.~T.}\ \bibnamefont {Boothroyd}}, \bibinfo
  {author} {\bibfnamefont {D.}~\bibnamefont {Prabhakaran}}, \bibinfo {author}
  {\bibfnamefont {A.}~\bibnamefont {Tanaka}}, \ and\ \bibinfo {author}
  {\bibfnamefont {M.}~\bibnamefont {Doerr}},\ }\href {\doibase
  10.1038/srep07003} {\bibfield  {journal} {\bibinfo  {journal} {Sci. Rep.}\
  }\textbf {\bibinfo {volume} {4}},\ \bibinfo {pages} {7003} (\bibinfo {year}
  {2014})}\BibitemShut {NoStop}%
\bibitem [{\citenamefont {Ikeda}\ \emph {et~al.}(2016)\citenamefont {Ikeda},
  \citenamefont {Nomura}, \citenamefont {Matsuda}, \citenamefont {Matsuo},
  \citenamefont {Kindo},\ and\ \citenamefont {Sato}}]{Ikeda}%
  \BibitemOpen
  \bibfield  {author} {\bibinfo {author} {\bibfnamefont {A.}~\bibnamefont
  {Ikeda}}, \bibinfo {author} {\bibfnamefont {T.}~\bibnamefont {Nomura}},
  \bibinfo {author} {\bibfnamefont {Y.~H.}\ \bibnamefont {Matsuda}}, \bibinfo
  {author} {\bibfnamefont {A.}~\bibnamefont {Matsuo}}, \bibinfo {author}
  {\bibfnamefont {K.}~\bibnamefont {Kindo}}, \ and\ \bibinfo {author}
  {\bibfnamefont {K.}~\bibnamefont {Sato}},\ }\href {\doibase
  10.1103/PhysRevB.93.220401} {\bibfield  {journal} {\bibinfo  {journal} {Phys.
  Rev. B}\ }\textbf {\bibinfo {volume} {93}},\ \bibinfo {pages} {220401(R)}
  (\bibinfo {year} {2016})}\BibitemShut {NoStop}%
\bibitem [{\citenamefont {Sotnikov}\ and\ \citenamefont
  {Kune\v{s}}(2016)}]{Sotnikov}%
  \BibitemOpen
  \bibfield  {author} {\bibinfo {author} {\bibfnamefont {A.}~\bibnamefont
  {Sotnikov}}\ and\ \bibinfo {author} {\bibfnamefont {J.}~\bibnamefont
  {Kune\v{s}}},\ }\href {\doibase 10.1038/srep30510} {\bibfield  {journal}
  {\bibinfo  {journal} {Sci. Rep.}\ }\textbf {\bibinfo {volume} {6}},\ \bibinfo
  {pages} {30510} (\bibinfo {year} {2016})}\BibitemShut {NoStop}%
\bibitem [{\citenamefont {Tatsuno}\ \emph {et~al.}(2016)\citenamefont
  {Tatsuno}, \citenamefont {Mizoguchi}, \citenamefont {Nasu}, \citenamefont
  {Naka},\ and\ \citenamefont {Ishihara}}]{Tatsuno}%
  \BibitemOpen
  \bibfield  {author} {\bibinfo {author} {\bibfnamefont {T.}~\bibnamefont
  {Tatsuno}}, \bibinfo {author} {\bibfnamefont {E.}~\bibnamefont {Mizoguchi}},
  \bibinfo {author} {\bibfnamefont {J.}~\bibnamefont {Nasu}}, \bibinfo {author}
  {\bibfnamefont {M.}~\bibnamefont {Naka}}, \ and\ \bibinfo {author}
  {\bibfnamefont {S.}~\bibnamefont {Ishihara}},\ }\href {\doibase
  10.7566/JPSJ.85.083706} {\bibfield  {journal} {\bibinfo  {journal} {J. Phys.
  Soc. Jpn.}\ }\textbf {\bibinfo {volume} {85}},\ \bibinfo {pages} {083706}
  (\bibinfo {year} {2016})}\BibitemShut {NoStop}%
\bibitem [{\citenamefont {Naito}\ \emph {et~al.}(2014)\citenamefont {Naito},
  \citenamefont {Fujishiro}, \citenamefont {Nishizaki}, \citenamefont
  {Kobayashi}, \citenamefont {Hejtm\'{a}nek}, \citenamefont {Kn\'{i}\v{z}ek},\
  and\ \citenamefont {Jir\'{a}k}}]{Naito2014}%
  \BibitemOpen
  \bibfield  {author} {\bibinfo {author} {\bibfnamefont {T.}~\bibnamefont
  {Naito}}, \bibinfo {author} {\bibfnamefont {H.}~\bibnamefont {Fujishiro}},
  \bibinfo {author} {\bibfnamefont {T.}~\bibnamefont {Nishizaki}}, \bibinfo
  {author} {\bibfnamefont {N.}~\bibnamefont {Kobayashi}}, \bibinfo {author}
  {\bibfnamefont {J.}~\bibnamefont {Hejtm\'{a}nek}}, \bibinfo {author}
  {\bibfnamefont {K.}~\bibnamefont {Kn\'{i}\v{z}ek}}, \ and\ \bibinfo {author}
  {\bibfnamefont {Z.}~\bibnamefont {Jir\'{a}k}},\ }\href {\doibase
  10.1063/1.4884435} {\bibfield  {journal} {\bibinfo  {journal} {J. Appl.
  Phys.}\ }\textbf {\bibinfo {volume} {115}},\ \bibinfo {pages} {233914}
  (\bibinfo {year} {2014})}\BibitemShut {NoStop}%
\bibitem [{\citenamefont {Lee}\ \emph {et~al.}(2015)\citenamefont {Lee},
  \citenamefont {Matsuda}, \citenamefont {Naito}, \citenamefont {Nakamura},\
  and\ \citenamefont {Takeyama}}]{Suyeon}%
  \BibitemOpen
  \bibfield  {author} {\bibinfo {author} {\bibfnamefont {S.}~\bibnamefont
  {Lee}}, \bibinfo {author} {\bibfnamefont {Y.~H.}\ \bibnamefont {Matsuda}},
  \bibinfo {author} {\bibfnamefont {T.}~\bibnamefont {Naito}}, \bibinfo
  {author} {\bibfnamefont {D.}~\bibnamefont {Nakamura}}, \ and\ \bibinfo
  {author} {\bibfnamefont {S.}~\bibnamefont {Takeyama}},\ }\href {\doibase
  10.7566/JPSJ.84.044703} {\bibfield  {journal} {\bibinfo  {journal} {J. Phys.
  Soc. Jpn.}\ }\textbf {\bibinfo {volume} {84}},\ \bibinfo {pages} {044703}
  (\bibinfo {year} {2015})}\BibitemShut {NoStop}%
\bibitem [{\citenamefont {Kimura}\ \emph {et~al.}(2005)\citenamefont {Kimura},
  \citenamefont {Narumi}, \citenamefont {Kindo}, \citenamefont {Nakano},\ and\
  \citenamefont {e.~Matsubayashi}}]{Kimura2005}%
  \BibitemOpen
  \bibfield  {author} {\bibinfo {author} {\bibfnamefont {S.}~\bibnamefont
  {Kimura}}, \bibinfo {author} {\bibfnamefont {Y.}~\bibnamefont {Narumi}},
  \bibinfo {author} {\bibfnamefont {K.}~\bibnamefont {Kindo}}, \bibinfo
  {author} {\bibfnamefont {M.}~\bibnamefont {Nakano}}, \ and\ \bibinfo {author}
  {\bibfnamefont {G.}~\bibnamefont {e.~Matsubayashi}},\ }\href {\doibase
  10.1103/PhysRevB.72.064448} {\bibfield  {journal} {\bibinfo  {journal} {Phys.
  Rev. B}\ }\textbf {\bibinfo {volume} {72}},\ \bibinfo {pages} {064448}
  (\bibinfo {year} {2005})}\BibitemShut {NoStop}%
\bibitem [{\citenamefont {Kimura}\ \emph {et~al.}(2008)\citenamefont {Kimura},
  \citenamefont {Maeda}, \citenamefont {Kashiwagi}, \citenamefont {Yamaguchi},
  \citenamefont {Hagiwara}, \citenamefont {Yoshida}, \citenamefont {Terasaki},\
  and\ \citenamefont {Kindo}}]{Kimura2008}%
  \BibitemOpen
  \bibfield  {author} {\bibinfo {author} {\bibfnamefont {S.}~\bibnamefont
  {Kimura}}, \bibinfo {author} {\bibfnamefont {Y.}~\bibnamefont {Maeda}},
  \bibinfo {author} {\bibfnamefont {T.}~\bibnamefont {Kashiwagi}}, \bibinfo
  {author} {\bibfnamefont {H.}~\bibnamefont {Yamaguchi}}, \bibinfo {author}
  {\bibfnamefont {M.}~\bibnamefont {Hagiwara}}, \bibinfo {author}
  {\bibfnamefont {S.}~\bibnamefont {Yoshida}}, \bibinfo {author} {\bibfnamefont
  {I.}~\bibnamefont {Terasaki}}, \ and\ \bibinfo {author} {\bibfnamefont
  {K.}~\bibnamefont {Kindo}},\ }\href {\doibase 10.1103/PhysRevB.78.180403}
  {\bibfield  {journal} {\bibinfo  {journal} {Phys. Rev. B}\ }\textbf {\bibinfo
  {volume} {78}},\ \bibinfo {pages} {180403} (\bibinfo {year}
  {2008})}\BibitemShut {NoStop}%
\bibitem [{\citenamefont {Takeyama}\ \emph {et~al.}(2012)\citenamefont
  {Takeyama}, \citenamefont {Sakakura}, \citenamefont {Matsuda}, \citenamefont
  {Miyata},\ and\ \citenamefont {Tokunaga}}]{Takeyama2012}%
  \BibitemOpen
  \bibfield  {author} {\bibinfo {author} {\bibfnamefont {S.}~\bibnamefont
  {Takeyama}}, \bibinfo {author} {\bibfnamefont {R.}~\bibnamefont {Sakakura}},
  \bibinfo {author} {\bibfnamefont {Y.~H.}\ \bibnamefont {Matsuda}}, \bibinfo
  {author} {\bibfnamefont {A.}~\bibnamefont {Miyata}}, \ and\ \bibinfo {author}
  {\bibfnamefont {M.}~\bibnamefont {Tokunaga}},\ }\href {\doibase
  10.1143/JPSJ.81.014702} {\bibfield  {journal} {\bibinfo  {journal} {J. Phys.
  Soc. Jpn.}\ }\textbf {\bibinfo {volume} {81}},\ \bibinfo {pages} {014702}
  (\bibinfo {year} {2012})}\BibitemShut {NoStop}%
\bibitem [{\citenamefont {Amaya}\ \emph {et~al.}(1989)\citenamefont {Amaya},
  \citenamefont {Takeyama}, \citenamefont {Nakagawa}, \citenamefont {Ishizuka},
  \citenamefont {Nakao}, \citenamefont {Sakakibara}, \citenamefont {Goto},
  \citenamefont {Miura}, \citenamefont {Ajiro},\ and\ \citenamefont
  {Kikuchi}}]{Amaya}%
  \BibitemOpen
  \bibfield  {author} {\bibinfo {author} {\bibfnamefont {K.}~\bibnamefont
  {Amaya}}, \bibinfo {author} {\bibfnamefont {S.}~\bibnamefont {Takeyama}},
  \bibinfo {author} {\bibfnamefont {T.}~\bibnamefont {Nakagawa}}, \bibinfo
  {author} {\bibfnamefont {M.}~\bibnamefont {Ishizuka}}, \bibinfo {author}
  {\bibfnamefont {K.}~\bibnamefont {Nakao}}, \bibinfo {author} {\bibfnamefont
  {T.}~\bibnamefont {Sakakibara}}, \bibinfo {author} {\bibfnamefont
  {T.}~\bibnamefont {Goto}}, \bibinfo {author} {\bibfnamefont {N.}~\bibnamefont
  {Miura}}, \bibinfo {author} {\bibfnamefont {Y.}~\bibnamefont {Ajiro}}, \ and\
  \bibinfo {author} {\bibfnamefont {H.}~\bibnamefont {Kikuchi}},\ }\href
  {\doibase 10.1016/0921-4526(89)90538-3} {\bibfield  {journal} {\bibinfo
  {journal} {Physica B}\ }\textbf {\bibinfo {volume} {155}},\ \bibinfo {pages}
  {396} (\bibinfo {year} {1989})}\BibitemShut {NoStop}%
\bibitem [{\citenamefont {Hejtm\'{a}nek}\ \emph {et~al.}(2013)\citenamefont
  {Hejtm\'{a}nek}, \citenamefont {Jir\'{a}k}, \citenamefont {Kaman},
  \citenamefont {Kn\'{i}\v{z}ek}, \citenamefont {\v{S}antav\'{a}},
  \citenamefont {Nitta}, \citenamefont {Naito},\ and\ \citenamefont
  {Fujishiro}}]{Hejtmanek2013}%
  \BibitemOpen
  \bibfield  {author} {\bibinfo {author} {\bibfnamefont {J.}~\bibnamefont
  {Hejtm\'{a}nek}}, \bibinfo {author} {\bibfnamefont {Z.}~\bibnamefont
  {Jir\'{a}k}}, \bibinfo {author} {\bibfnamefont {O.}~\bibnamefont {Kaman}},
  \bibinfo {author} {\bibfnamefont {K.}~\bibnamefont {Kn\'{i}\v{z}ek}},
  \bibinfo {author} {\bibfnamefont {E.}~\bibnamefont {\v{S}antav\'{a}}},
  \bibinfo {author} {\bibfnamefont {K.}~\bibnamefont {Nitta}}, \bibinfo
  {author} {\bibfnamefont {T.}~\bibnamefont {Naito}}, \ and\ \bibinfo {author}
  {\bibfnamefont {H.}~\bibnamefont {Fujishiro}},\ }\href {\doibase
  10.1140/epjb/e2013-30653-y} {\bibfield  {journal} {\bibinfo  {journal} {Eur.
  Phys. J. B}\ }\textbf {\bibinfo {volume} {86}},\ \bibinfo {pages} {305}
  (\bibinfo {year} {2013})}\BibitemShut {NoStop}%
\bibitem [{\citenamefont {Nomura}\ \emph {et~al.}(2016)\citenamefont {Nomura},
  \citenamefont {Matsuda}, \citenamefont {Takeyama},\ and\ \citenamefont
  {Kobayashi}}]{Nomura}%
  \BibitemOpen
  \bibfield  {author} {\bibinfo {author} {\bibfnamefont {T.}~\bibnamefont
  {Nomura}}, \bibinfo {author} {\bibfnamefont {Y.~H.}\ \bibnamefont {Matsuda}},
  \bibinfo {author} {\bibfnamefont {S.}~\bibnamefont {Takeyama}}, \ and\
  \bibinfo {author} {\bibfnamefont {T.~C.}\ \bibnamefont {Kobayashi}},\ }\href
  {\doibase 10.7566/JPSJ.85.094601} {\bibfield  {journal} {\bibinfo  {journal}
  {J. Phys. Soc. Jpn.}\ }\textbf {\bibinfo {volume} {85}},\ \bibinfo {pages}
  {094601} (\bibinfo {year} {2016})}\BibitemShut {NoStop}%
\bibitem [{\citenamefont {Biernacki}\ and\ \citenamefont
  {Clerjaud}(2005)}]{Biernacki2005}%
  \BibitemOpen
  \bibfield  {author} {\bibinfo {author} {\bibfnamefont {S.~W.}\ \bibnamefont
  {Biernacki}}\ and\ \bibinfo {author} {\bibfnamefont {B.}~\bibnamefont
  {Clerjaud}},\ }\href {\doibase 10.1103/PhysRevB.72.024406} {\bibfield
  {journal} {\bibinfo  {journal} {Phys. Rev. B}\ }\textbf {\bibinfo {volume}
  {72}},\ \bibinfo {pages} {024406} (\bibinfo {year} {2005})}\BibitemShut
  {NoStop}%
\bibitem [{\citenamefont {Itoh}\ \emph {et~al.}(1994)\citenamefont {Itoh},
  \citenamefont {Natori}, \citenamefont {Kubota},\ and\ \citenamefont
  {Motoya}}]{Itoh}%
  \BibitemOpen
  \bibfield  {author} {\bibinfo {author} {\bibfnamefont {M.}~\bibnamefont
  {Itoh}}, \bibinfo {author} {\bibfnamefont {I.}~\bibnamefont {Natori}},
  \bibinfo {author} {\bibfnamefont {S.}~\bibnamefont {Kubota}}, \ and\ \bibinfo
  {author} {\bibfnamefont {K.}~\bibnamefont {Motoya}},\ }\href {\doibase
  10.1143/JPSJ.63.1486} {\bibfield  {journal} {\bibinfo  {journal} {J. Phys.
  Soc. Jpn.}\ }\textbf {\bibinfo {volume} {63}},\ \bibinfo {pages} {1486}
  (\bibinfo {year} {1994})}\BibitemShut {NoStop}%
\bibitem [{\citenamefont {Saitoh}\ \emph
  {et~al.}(1997{\natexlab{c}})\citenamefont {Saitoh}, \citenamefont {Sekiyama},
  \citenamefont {Kobayashi}, \citenamefont {Mizokawa}, \citenamefont
  {Fujimori}, \citenamefont {Sarma}, \citenamefont {Takeda},\ and\
  \citenamefont {Takano}}]{Saitoh2}%
  \BibitemOpen
  \bibfield  {author} {\bibinfo {author} {\bibfnamefont {T.}~\bibnamefont
  {Saitoh}}, \bibinfo {author} {\bibfnamefont {A.}~\bibnamefont {Sekiyama}},
  \bibinfo {author} {\bibfnamefont {K.}~\bibnamefont {Kobayashi}}, \bibinfo
  {author} {\bibfnamefont {T.}~\bibnamefont {Mizokawa}}, \bibinfo {author}
  {\bibfnamefont {A.}~\bibnamefont {Fujimori}}, \bibinfo {author}
  {\bibfnamefont {D.~D.}\ \bibnamefont {Sarma}}, \bibinfo {author}
  {\bibfnamefont {Y.}~\bibnamefont {Takeda}}, \ and\ \bibinfo {author}
  {\bibfnamefont {M.}~\bibnamefont {Takano}},\ }\href {\doibase
  10.1103/PhysRevB.56.8836} {\bibfield  {journal} {\bibinfo  {journal} {Phys.
  Rev. B}\ }\textbf {\bibinfo {volume} {56}},\ \bibinfo {pages} {8836}
  (\bibinfo {year} {1997}{\natexlab{c}})}\BibitemShut {NoStop}%
\bibitem [{\citenamefont {Okamoto}\ \emph {et~al.}(2000)\citenamefont
  {Okamoto}, \citenamefont {Miyauchi}, \citenamefont {Sekine}, \citenamefont
  {Shidara}, \citenamefont {Koide}, \citenamefont {Amemiya}, \citenamefont
  {Fujimori}, \citenamefont {Saitoh}, \citenamefont {Tanaka}, \citenamefont
  {Takeda},\ and\ \citenamefont {Takano}}]{Okamoto}%
  \BibitemOpen
  \bibfield  {author} {\bibinfo {author} {\bibfnamefont {J.}~\bibnamefont
  {Okamoto}}, \bibinfo {author} {\bibfnamefont {H.}~\bibnamefont {Miyauchi}},
  \bibinfo {author} {\bibfnamefont {T.}~\bibnamefont {Sekine}}, \bibinfo
  {author} {\bibfnamefont {T.}~\bibnamefont {Shidara}}, \bibinfo {author}
  {\bibfnamefont {T.}~\bibnamefont {Koide}}, \bibinfo {author} {\bibfnamefont
  {K.}~\bibnamefont {Amemiya}}, \bibinfo {author} {\bibfnamefont
  {A.}~\bibnamefont {Fujimori}}, \bibinfo {author} {\bibfnamefont
  {T.}~\bibnamefont {Saitoh}}, \bibinfo {author} {\bibfnamefont
  {A.}~\bibnamefont {Tanaka}}, \bibinfo {author} {\bibfnamefont
  {Y.}~\bibnamefont {Takeda}}, \ and\ \bibinfo {author} {\bibfnamefont
  {M.}~\bibnamefont {Takano}},\ }\href {\doibase 10.1103/PhysRevB.62.4455}
  {\bibfield  {journal} {\bibinfo  {journal} {Phys. Rev. B}\ }\textbf {\bibinfo
  {volume} {62}},\ \bibinfo {pages} {4455} (\bibinfo {year}
  {2000})}\BibitemShut {NoStop}%
\bibitem [{\citenamefont {Nov\'{a}k}\ \emph {et~al.}(2013)\citenamefont
  {Nov\'{a}k}, \citenamefont {Kn\'{i}\v{v}ek}, \citenamefont {Mary\v{s}ko},
  \citenamefont {Jir\'{a}k},\ and\ \citenamefont {Kune\v{s}}}]{Novak}%
  \BibitemOpen
  \bibfield  {author} {\bibinfo {author} {\bibfnamefont {P.}~\bibnamefont
  {Nov\'{a}k}}, \bibinfo {author} {\bibfnamefont {K.}~\bibnamefont
  {Kn\'{i}\v{v}ek}}, \bibinfo {author} {\bibfnamefont {M.}~\bibnamefont
  {Mary\v{s}ko}}, \bibinfo {author} {\bibfnamefont {Z.}~\bibnamefont
  {Jir\'{a}k}}, \ and\ \bibinfo {author} {\bibfnamefont {J.}~\bibnamefont
  {Kune\v{s}}},\ }\href {\doibase 10.1088/0953-8984/25/44/446001} {\bibfield
  {journal} {\bibinfo  {journal} {J. Phys.: Cond. Mat.}\ }\textbf {\bibinfo
  {volume} {25}},\ \bibinfo {pages} {446001} (\bibinfo {year}
  {2013})}\BibitemShut {NoStop}%
\bibitem [{\citenamefont {Jir\'{a}k}\ \emph {et~al.}(2013)\citenamefont
  {Jir\'{a}k}, \citenamefont {Hejtm\'{a}nek}, \citenamefont {Kn\'{i}\v{z}ek},
  \citenamefont {Mary\v{s}ko}, \citenamefont {Nov\'{a}k}, \citenamefont
  {\v{S}antav\'{a}}, \citenamefont {Naito},\ and\ \citenamefont
  {Fujishiro}}]{Jirak}%
  \BibitemOpen
  \bibfield  {author} {\bibinfo {author} {\bibfnamefont {Z.}~\bibnamefont
  {Jir\'{a}k}}, \bibinfo {author} {\bibfnamefont {J.}~\bibnamefont
  {Hejtm\'{a}nek}}, \bibinfo {author} {\bibfnamefont {K.}~\bibnamefont
  {Kn\'{i}\v{z}ek}}, \bibinfo {author} {\bibfnamefont {M.}~\bibnamefont
  {Mary\v{s}ko}}, \bibinfo {author} {\bibfnamefont {P.}~\bibnamefont
  {Nov\'{a}k}}, \bibinfo {author} {\bibfnamefont {E.}~\bibnamefont
  {\v{S}antav\'{a}}}, \bibinfo {author} {\bibfnamefont {T.}~\bibnamefont
  {Naito}}, \ and\ \bibinfo {author} {\bibfnamefont {H.}~\bibnamefont
  {Fujishiro}},\ }\href {\doibase 10.1088/0953-8984/25/21/216006} {\bibfield
  {journal} {\bibinfo  {journal} {J. Phys.: Cond. Mat.}\ }\textbf {\bibinfo
  {volume} {25}},\ \bibinfo {pages} {216006} (\bibinfo {year}
  {2013})}\BibitemShut {NoStop}%
\bibitem [{\citenamefont {Hejtm\'{a}nek}\ \emph {et~al.}(2010)\citenamefont
  {Hejtm\'{a}nek}, \citenamefont {\v{S}antav\'{a}}, \citenamefont
  {Kn\'{i}\v{z}ek}, \citenamefont {Mary\v{s}ko}, \citenamefont {Jir\'{a}k},
  \citenamefont {Naito}, \citenamefont {Sasaki},\ and\ \citenamefont
  {Fujishiro}}]{Hejtmanek2010}%
  \BibitemOpen
  \bibfield  {author} {\bibinfo {author} {\bibfnamefont {J.}~\bibnamefont
  {Hejtm\'{a}nek}}, \bibinfo {author} {\bibfnamefont {E.}~\bibnamefont
  {\v{S}antav\'{a}}}, \bibinfo {author} {\bibnamefont {Kn\'{i}\v{z}ek}},
  \bibinfo {author} {\bibfnamefont {M.}~\bibnamefont {Mary\v{s}ko}}, \bibinfo
  {author} {\bibfnamefont {Z.}~\bibnamefont {Jir\'{a}k}}, \bibinfo {author}
  {\bibfnamefont {T.}~\bibnamefont {Naito}}, \bibinfo {author} {\bibfnamefont
  {H.}~\bibnamefont {Sasaki}}, \ and\ \bibinfo {author} {\bibfnamefont
  {H.}~\bibnamefont {Fujishiro}},\ }\href {\doibase 10.1103/PhysRevB.82.165107}
  {\bibfield  {journal} {\bibinfo  {journal} {Phys. Rev. B}\ }\textbf {\bibinfo
  {volume} {82}},\ \bibinfo {pages} {165107} (\bibinfo {year}
  {2010})}\BibitemShut {NoStop}%
\bibitem [{\citenamefont {Yamaguchi}\ \emph {et~al.}(1995)\citenamefont
  {Yamaguchi}, \citenamefont {Taniguchi}, \citenamefont {Takagi}, \citenamefont
  {Arima},\ and\ \citenamefont {Tokura}}]{Yamaguchi1995}%
  \BibitemOpen
  \bibfield  {author} {\bibinfo {author} {\bibfnamefont {S.}~\bibnamefont
  {Yamaguchi}}, \bibinfo {author} {\bibfnamefont {H.}~\bibnamefont
  {Taniguchi}}, \bibinfo {author} {\bibfnamefont {H.}~\bibnamefont {Takagi}},
  \bibinfo {author} {\bibfnamefont {T.}~\bibnamefont {Arima}}, \ and\ \bibinfo
  {author} {\bibfnamefont {Y.}~\bibnamefont {Tokura}},\ }\href {\doibase
  10.1143/JPSJ.64.1885} {\bibfield  {journal} {\bibinfo  {journal} {J. Phys.
  Soc. Jpn.}\ }\textbf {\bibinfo {volume} {64}},\ \bibinfo {pages} {1885}
  (\bibinfo {year} {1995})}\BibitemShut {NoStop}%
\bibitem [{\citenamefont {Yamaguchi}\ \emph {et~al.}(1996)\citenamefont
  {Yamaguchi}, \citenamefont {Okimoto}, \citenamefont {Taniguchi},\ and\
  \citenamefont {Tokura}}]{Yamaguchi1996}%
  \BibitemOpen
  \bibfield  {author} {\bibinfo {author} {\bibfnamefont {S.}~\bibnamefont
  {Yamaguchi}}, \bibinfo {author} {\bibfnamefont {Y.}~\bibnamefont {Okimoto}},
  \bibinfo {author} {\bibfnamefont {H.}~\bibnamefont {Taniguchi}}, \ and\
  \bibinfo {author} {\bibfnamefont {Y.}~\bibnamefont {Tokura}},\ }\href
  {\doibase 10.1103/PhysRevB.53.R2926} {\bibfield  {journal} {\bibinfo
  {journal} {Phys. Rev. B}\ }\textbf {\bibinfo {volume} {53}},\ \bibinfo
  {pages} {R2926} (\bibinfo {year} {1996})}\BibitemShut {NoStop}%
\bibitem [{\citenamefont {Mary\v{s}ko}\ \emph {et~al.}(2011)\citenamefont
  {Mary\v{s}ko}, \citenamefont {Jir\'{a}k}, \citenamefont {Kn\'{i}\v{z}ek},
  \citenamefont {Nov\'{a}k}, \citenamefont {Hejtm\'{a}nek}, \citenamefont
  {Naito}, \citenamefont {Sasaki},\ and\ \citenamefont {Fujishiro}}]{Marysko}%
  \BibitemOpen
  \bibfield  {author} {\bibinfo {author} {\bibfnamefont {M.}~\bibnamefont
  {Mary\v{s}ko}}, \bibinfo {author} {\bibfnamefont {Z.}~\bibnamefont
  {Jir\'{a}k}}, \bibinfo {author} {\bibfnamefont {K.}~\bibnamefont
  {Kn\'{i}\v{z}ek}}, \bibinfo {author} {\bibfnamefont {P.}~\bibnamefont
  {Nov\'{a}k}}, \bibinfo {author} {\bibfnamefont {J.}~\bibnamefont
  {Hejtm\'{a}nek}}, \bibinfo {author} {\bibfnamefont {T.}~\bibnamefont
  {Naito}}, \bibinfo {author} {\bibfnamefont {H.}~\bibnamefont {Sasaki}}, \
  and\ \bibinfo {author} {\bibfnamefont {H.}~\bibnamefont {Fujishiro}},\ }\href
  {\doibase 10.1063/1.3559485} {\bibfield  {journal} {\bibinfo  {journal} {J.
  Appl. Phys.}\ }\textbf {\bibinfo {volume} {109}},\ \bibinfo {pages} {07E127}
  (\bibinfo {year} {2011})}\BibitemShut {NoStop}%
\bibitem [{\citenamefont {Hejtm\'{a}nek}\ \emph {et~al.}()\citenamefont
  {Hejtm\'{a}nek}, \citenamefont {Jir\'{a}k}, \citenamefont {Kaman},
  \citenamefont {Kn\'{i}\v{z}ek}, \citenamefont {\v{S}antav\'{a}},
  \citenamefont {Nitta}, \citenamefont {Naito},\ and\ \citenamefont
  {Fujishiro}}]{HejtmanekPrivate}%
  \BibitemOpen
  \bibfield  {author} {\bibinfo {author} {\bibfnamefont {J.}~\bibnamefont
  {Hejtm\'{a}nek}}, \bibinfo {author} {\bibfnamefont {Z.}~\bibnamefont
  {Jir\'{a}k}}, \bibinfo {author} {\bibfnamefont {O.}~\bibnamefont {Kaman}},
  \bibinfo {author} {\bibfnamefont {K.}~\bibnamefont {Kn\'{i}\v{z}ek}},
  \bibinfo {author} {\bibfnamefont {E.}~\bibnamefont {\v{S}antav\'{a}}},
  \bibinfo {author} {\bibfnamefont {K.}~\bibnamefont {Nitta}}, \bibinfo
  {author} {\bibfnamefont {T.}~\bibnamefont {Naito}}, \ and\ \bibinfo {author}
  {\bibfnamefont {H.}~\bibnamefont {Fujishiro}},\ }\href@noop {} {}\bibinfo
  {howpublished} {(private communication)}\BibitemShut {NoStop}%
\bibitem [{\citenamefont {Moritomo}\ \emph {et~al.}(1995)\citenamefont
  {Moritomo}, \citenamefont {Asamitsu},\ and\ \citenamefont
  {Tokura}}]{Morimoto1995}%
  \BibitemOpen
  \bibfield  {author} {\bibinfo {author} {\bibfnamefont {Y.}~\bibnamefont
  {Moritomo}}, \bibinfo {author} {\bibfnamefont {A.}~\bibnamefont {Asamitsu}},
  \ and\ \bibinfo {author} {\bibfnamefont {Y.}~\bibnamefont {Tokura}},\ }\href
  {\doibase 10.1103/PhysRevB.51.16491} {\bibfield  {journal} {\bibinfo
  {journal} {Phys. Rev. B}\ }\textbf {\bibinfo {volume} {51}},\ \bibinfo
  {pages} {16491} (\bibinfo {year} {1995})}\BibitemShut {NoStop}%
\bibitem [{\citenamefont {Moritomo}\ \emph {et~al.}(1997)\citenamefont
  {Moritomo}, \citenamefont {Kuwahara}, \citenamefont {Tomioka},\ and\
  \citenamefont {Tokura}}]{Morimoto1997}%
  \BibitemOpen
  \bibfield  {author} {\bibinfo {author} {\bibfnamefont {Y.}~\bibnamefont
  {Moritomo}}, \bibinfo {author} {\bibfnamefont {H.}~\bibnamefont {Kuwahara}},
  \bibinfo {author} {\bibfnamefont {Y.}~\bibnamefont {Tomioka}}, \ and\
  \bibinfo {author} {\bibfnamefont {Y.}~\bibnamefont {Tokura}},\ }\href
  {\doibase 10.1103/PhysRevB.55.7549} {\bibfield  {journal} {\bibinfo
  {journal} {Phys. Rev. B}\ }\textbf {\bibinfo {volume} {55}},\ \bibinfo
  {pages} {7549} (\bibinfo {year} {1997})}\BibitemShut {NoStop}%
\bibitem [{\citenamefont {Lengsdorf}\ \emph {et~al.}(2004)\citenamefont
  {Lengsdorf}, \citenamefont {Ait-Tahar}, \citenamefont {Saxena}, \citenamefont
  {Ellerby}, \citenamefont {Khomskii}, \citenamefont {Micklitz}, \citenamefont
  {Lorenz},\ and\ \citenamefont {Abd-Elmeguid}}]{Lengsdorf}%
  \BibitemOpen
  \bibfield  {author} {\bibinfo {author} {\bibfnamefont {R.}~\bibnamefont
  {Lengsdorf}}, \bibinfo {author} {\bibfnamefont {M.}~\bibnamefont
  {Ait-Tahar}}, \bibinfo {author} {\bibfnamefont {S.~S.}\ \bibnamefont
  {Saxena}}, \bibinfo {author} {\bibfnamefont {M.}~\bibnamefont {Ellerby}},
  \bibinfo {author} {\bibfnamefont {D.~I.}\ \bibnamefont {Khomskii}}, \bibinfo
  {author} {\bibfnamefont {H.}~\bibnamefont {Micklitz}}, \bibinfo {author}
  {\bibfnamefont {T.}~\bibnamefont {Lorenz}}, \ and\ \bibinfo {author}
  {\bibfnamefont {M.~M.}\ \bibnamefont {Abd-Elmeguid}},\ }\href {\doibase
  10.1103/PhysRevB.69.140403} {\bibfield  {journal} {\bibinfo  {journal} {Phys.
  Rev. B}\ }\textbf {\bibinfo {volume} {69}},\ \bibinfo {pages} {140403}
  (\bibinfo {year} {2004})}\BibitemShut {NoStop}%
\bibitem [{\citenamefont {Fita}\ \emph {et~al.}(2005)\citenamefont {Fita},
  \citenamefont {Szymczak}, \citenamefont {Puzniak}, \citenamefont
  {Troyanchuk}, \citenamefont {Fink-Finowicki}, \citenamefont {Mukovskii},
  \citenamefont {Varyukhin},\ and\ \citenamefont {Szymczak}}]{Fita}%
  \BibitemOpen
  \bibfield  {author} {\bibinfo {author} {\bibfnamefont {I.}~\bibnamefont
  {Fita}}, \bibinfo {author} {\bibfnamefont {R.}~\bibnamefont {Szymczak}},
  \bibinfo {author} {\bibfnamefont {R.}~\bibnamefont {Puzniak}}, \bibinfo
  {author} {\bibfnamefont {I.~O.}\ \bibnamefont {Troyanchuk}}, \bibinfo
  {author} {\bibfnamefont {J.}~\bibnamefont {Fink-Finowicki}}, \bibinfo
  {author} {\bibfnamefont {Y.~M.}\ \bibnamefont {Mukovskii}}, \bibinfo {author}
  {\bibfnamefont {V.~N.}\ \bibnamefont {Varyukhin}}, \ and\ \bibinfo {author}
  {\bibfnamefont {H.}~\bibnamefont {Szymczak}},\ }\href {\doibase
  10.1103/PhysRevB.71.214404} {\bibfield  {journal} {\bibinfo  {journal} {Phys.
  Rev. B}\ }\textbf {\bibinfo {volume} {71}},\ \bibinfo {pages} {214404}
  (\bibinfo {year} {2005})}\BibitemShut {NoStop}%
\bibitem [{\citenamefont {Chesnut}(1964)}]{Chesnut}%
  \BibitemOpen
  \bibfield  {author} {\bibinfo {author} {\bibfnamefont {D.~B.}\ \bibnamefont
  {Chesnut}},\ }\href {\doibase 10.1063/1.1725127} {\bibfield  {journal}
  {\bibinfo  {journal} {J. Chem. Phys.}\ }\textbf {\bibinfo {volume} {40}},\
  \bibinfo {pages} {405} (\bibinfo {year} {1964})}\BibitemShut {NoStop}%
\end{thebibliography}%

\appendix

\section{A field-induced spin crossover model}

The spin crossover model of single ion formulated by Biernacki \cite{Biernacki2005} is simplified and summarized below, where spin crossover between LS and IS is induced by temperature or magnetic field, whose energy diagram is schematically shown in the inset of Fig. \ref{sim} (a). 
The inter ion interactions is taken into account under the mean field approximation. The inter ion interactions are effectively expressed by the elastic lattice model \cite{Chesnut}. In the elastic lattice model, the energy of LS and IS state are expressed as,
\begin{eqnarray}
E_{\mathrm{LS}}&&=\epsilon n^{2},\\
E_{\mathrm{IS}}&&=\Delta-g\mu_{\mathrm{B}}BS_{z}+\epsilon (n-1)^{2},
\end{eqnarray}
where  $k$, $\beta$, $g$, $\mu_{\mathrm{B}}$, $\Delta$, $S_{z}$ and $\epsilon$ are Boltzmann constant, $(kT)^{-1}$, g-factor, Bohr magneton, crystal field gap, spin angular momentum in $z$ direction and elastic constant of lattice, respectively.
$n$ is the increment of the lattice constant $a$ normalized as $n=\Delta a/(a_{\mathrm{IS}}-a_{\mathrm{LS}}$), where $a_{\mathrm{LS}}$ and $a_{\mathrm{IS}}$ are the lattice constant when all Co ion is in LS state and IS state, and $\Delta a = a-a_{\mathrm{LS}}$, respectively.
$n=0$ and 1 correspond to the state of lattice with uniformly occupied LS state and IS state, respectively.
According to Ref. \cite{Biernacki2005}, we can regard that $\epsilon$ is proportional to the inter ion interaction for the LS-LS or IS-IS pair.
We in this model note that $n$ is also regarded as the occupancy of the IS state \cite{Chesnut, Biernacki2005}.
Specifically, when $\epsilon>0$, the effective interaction for LS-LS and IS-IS pair are attractive and that for IS-LS pair is repulsive, giving rising to the first order spin state transition at finite temperature.
Thus, the partition function per single particle and $n$ can be defined as, 
\begin{equation}
Z=\exp\left(-\beta E_{\mathrm{LS}}\right)+\sum_{S_{z}=1, 0, -1}\exp\left(-\beta E_{\mathrm{IS}}\right),
\end{equation}
and
\begin{equation}
n=\frac{\sum_{S_{z}=1, 0, -1}\exp(-\beta E_{\mathrm{IS}})}{Z},
\end{equation}
, respectively. The one particle free energy is obtained as
$F=-\beta^{-1}\ln Z$.
Then, we obtain the equilibrium value of $n$ by obtaining the minimum of $F$ as
\begin{equation}
\frac{\partial F}{\partial n}=0,
\label{eqn}
\end{equation}
with the condition of $\partial^{2} F/\partial n^{2}>0$. The transition fields as a function of temperature $B_{\mathrm{C}}(T)$ can be obtained by numerically solving Eq. (\ref{eqn}) with the condition of $n=1/2$ as \cite{Biernacki2005}
\begin{equation}
T_{\mathrm{C}}=\frac{\Delta'}{k \ln P}
\label{tc1}
\end{equation}
where $\Delta'=\Delta-\epsilon$, and
\begin{equation}
P=\frac{\sinh \{\beta(2s+1)g\mu_{\mathrm{B}}B_{\mathrm{C}}/2\}}{\sinh(\beta g\mu_{\mathrm{B}}B_{\mathrm{C}}/2)},
\label{tc2}
\end{equation}
where $s$ takes 1 for IS. At $B=0$ T, Eq. (\ref{tc1}) and (\ref{tc2}) reduce to the form $T_{\mathrm{C}}=\Delta' / \{k\ln(2s+1)\}$. At $T=0$ K, on the other hand, Eq. (\ref{tc1}) and (\ref{tc2}) reduce to the form $B_{\mathrm{C}}=\Delta' / (g\mu_{\mathrm{B}}s)$. Both forms are readily understandable and useful to estimate $\Delta'$ and $g$.

\end{document}